\documentclass[article,amsmath,11pt,amssymb,floatfix,superscriptaddress,
showpagenumber,showpacs,nofootinbib]{revtex4}
\usepackage{graphicx}
\usepackage{epsfig}
\usepackage{bm}
\usepackage{amsfonts}
\usepackage{xcolor}
\usepackage{caption,subcaption}
\usepackage{amsmath}
\usepackage{tabularx}
\usepackage{caption}
\input epsf
\begin{document}
\title{Reconstruction of Tsallis Holographic Dark Energy via Modified Non-Metric Gravity: An $f(Q,C)$ Approach}

\author{Sanjeeda Sultana \footnote{Email: sanjeeda.sultana0401@gmail.com; sanjeeda.sultana1@s.amity.edu}}
\author{Surajit Chattopadhyay \footnote{Email: schattopadhyay1@kol.amity.edu; surajitchatto@outlook.com}}
\affiliation{ Department of Mathematics, Amity University Kolkata,\\
Major Arterial Road, Action Area II, Rajarhat, Newtown, Kolkata 700135,
India.}

\date{\today}

\newpage
\begin{abstract}
In the current research, we have reported the Tsallis Holographic Dark Energy (THDE) (\textit{JCAP}, 2018(12), p.012.) model reconstructed within the framework of $f(Q, C)$ gravity (\textit{JCAP}, 2024(03), p.050.), combining entropy-based dark energy models with geometrically motivated modified gravity to explain late-time cosmic acceleration. The reconstructed model is found to exhibit significant sensitivity to the parameter space $(H_0, a_0, n, \delta, \zeta,r_d)$ and the initial conditions. The evolution of the equation of state and deceleration parameters is found to be highly dependent on these parameters. A comprehensive Markov Chain Monte Carlo analysis using observational datasets comprising {CC+Pantheon$^{+}$+DESI DR2} was performed, yielding best-fit values that demonstrate strong consistency with observational data, which is further validated for its consistency through the computation of the age of the Universe. The evolution of the jerk and snap parameters is examined and compared with the $\Lambda$CDM prediction. Statefinder diagnostics, through the evolutionary trajectories of the pairs $(r, s)$ and $(r, q)$ are derived and indicate that the model passes through the $\Lambda$CDM fixed point and the physical viability of the model is further consolidated through analysis of the four energy conditions.
\end{abstract}


\maketitle

\section{Introduction}
{For the last couple of decades, the accelerated expansion of the Universe has emerged as one of the most exciting area of research in modern cosmology. This late-time acceleration is primarily explained by introducing a mysterious component known as dark energy \cite{riess1998observational, perlmutter1999measurements}, which is characterized be a negative equation of state parameter.The mysterious exotic dark energy has compelled the cosmologists to propose different candidates for dark energy and an abundance of literatures exploring various phenomenological models of dark energy are available in the literature \cite{saridakis2020,saridakis2018rgb,saridakis2008,copeland2006,nojiri2004,bamba2012}. One of the simplest explanations is the cosmological constant, first introduced by Einstein, which represents a constant energy density filling space homogeneously \cite{weinberg1989cosmological}. Another important model involving a slowly evolving scalar field is quintessence having some important dynamical properties \cite{ratra1988cosmological}. Other alternatives include the Chaplygin gas model, which attempts to unify dark energy and dark matter under a single framework \cite{kamenshchik2001alternative}, and phantom energy, where the equation of state parameter falls below \(-1\) and can lead to a Big Rip scenario \cite{caldwell2002phantom}.  The holographic dark energy (HDE) model has attracted interest for its relationship with quantum gravity principles \cite{li2004model}. An interesting generalization of HDE is the Tsallis holographic dark energy (THDE), which is based on the non-extensive Tsallis entropy formalism \cite{tsallis1988possible, tavayef2018tsallis}. We may recollect that the HDE model is based on the holographic principle, which acts as a connective the energy content of a system to its boundary area rather than its volume \cite{cohen1999effective}. However, the standard HDE model uses the Bekenstein-Hawking entropy.  This entropy assumes the usual additive property. Motivated by the idea that entropy could be non-additive in gravitational systems, Tsallis and Cirto proposed a generalized entropy formalism known as Tsallis entropy \cite{tsallis2013black}. On the basis of this proposal, the THDE model was introduced, where the energy density is modified according to the non-additive entropy principle \cite{D’Agostino}. In this framework of THDE, the dark energy density has a dependence on a power of the infrared (IR) cutoff with a power of nonextensivity parameter $\delta$. When $\delta = 1$, we get back the standard HDE. The THDE model provides more flexibility in describing the evolution of the Universe and can address some limitations of the standard HDE, such as better fitting to observational data and richer dynamical behavior. It has been studied in various cosmological scenarios including flat and non-flat Universes, and even in modified gravity frameworks \cite{tavayef2018tsallis}. This makes THDE a promising candidate for explaining dark energy from a thermodynamical perspective. } 

{Modified gravity theories, on the other hand, offer an alternative powerful method to understanding cosmic acceleration that does not involve dark energy. One such theory is \(f(Q,C) \) \cite{de2024non,SS,SS1}, which generalizes symmetric teleparallel gravity by introducing both the non-metricity scalar \(Q \) \cite{jimenez2020born} and a boundary term \(C \). Our goal is to reconstruct the THDE model using \(f(Q, C) \) gravity and investigate its cosmological consequences. At this juncture, let us first have a brief literature review of modified theories of gravity. Modified gravity is an important area of cosmology as it provides us with  alternative explanations for acceleration of the Universe, structure formation, and some clues towards a quantum theory of gravity. One of the earliest and most studied modifications is $f(R)$ gravity, where the Ricci scalar $R$ in the Einstein-Hilbert action is replaced by a general function $f(R)$ \cite{nojiri2006modified, sotiriou2010f}. This approach can naturally explain late-time acceleration without introducing a cosmological constant. Another interesting extension is $f(T)$ gravity, based on the teleparallel equivalent of GR, where $T$ is the torsion scalar instead of the curvature scalar \cite{cai2016f}. $f(T)$ theories have attracted attention due to their second-order field equations, which are simpler compared to $f(R)$ gravity. Recently, $f(Q)$ gravity has been proposed, where $Q$ is the non-metricity scalar associated with the gravitational interaction \cite{jimenez2018teleparallel}. $f(Q)$ gravity offers an alternative geometrical framework without curvature or torsion, and has shown potential in describing cosmic acceleration and alleviating cosmological tensions. As candidates of modified gravity, further generalizations have been proposed, such as $f(R,T)$ \cite{harko2011f}, $f(Q,B)$ \cite{capozziello2023}. These models involve non-minimal matter-geometry coupling. Furthermore,  they lead to interesting phenomenology in cosmological as well as astrophysical contexts. Moreover, scalar-tensor theories like Brans-Dicke theory \cite{brans1961mach} and Horndeski gravity \cite{horndeski1974second} represent important classes of modified gravity where a scalar field interacts with gravity, offering rich dynamical behaviors while remaining consistent with observational constraints.}

Let us now have a brief discussion on the holographic reconstruction of modified gravity theories. The holographic principle, initially proposed in the context of black hole physics, suggests that all the information within a volume of space can be described by degrees of freedom on its boundary \cite{tHooft1993dimensional}. In cosmology, this idea has been used to formulate the Holographic Dark Energy (HDE) model, where the dark energy density is related to the size of the cosmic horizon \cite{li2004model}. Motivated by the success of HDE in explaining the late-time acceleration, researchers have attempted to reconstruct modified gravity models, such as $f(R)$ and $f(Q)$ gravities, directly from holographic considerations \cite{nojiri2017modified}.
In holographic reconstruction, the evolution of the Hubble parameter or cosmological quantities inferred from HDE is used to determine the functional form of the gravitational action. For example, by assuming a specific behavior for the HDE density, it is possible to reconstruct an $f(R)$ function that mimics the desired cosmological dynamics \cite{Khodam-Mohammadi}. Similarly, the recent development of $f(Q)$ gravity, based on the non-metricity scalar $Q$, has opened new pathways to perform holographic reconstruction within a different geometric framework \cite{jimenez2018teleparallel}. Holographic reconstruction \cite{AP} provides a powerful link between quantum gravity ideas and cosmological observations, offering a novel route to building viable modified gravity models that are consistent with the holographic principle. Cosmological implications and thermodynamics of some reconstructed modified gravity models were explored for the thermodynamic behavior and cosmological impact of modified gravity models by \cite{azhar2018, jawad2020}. 

In order to consistently apply holographic principles and entropy relations to the Universe as a whole that constitutes a non-extensive thermodynamic system, it is essential to adopt a generalized definition of entropy. Specifically, employing the Tsallis entropy framework provides a more appropriate description of the Universe's horizon entropy. In the context of HDE, derived from the inequality $\rho_{\text{DE}} L^4 \leq S$ with $S \propto A \propto L^2$~\cite{saridakis2018tsallis1}, a consistent formulation arises when Tsallis entropy is inserted into this relation. This leads to a modified energy density expression that better encapsulates the thermodynamic properties of the cosmic horizon under non-extensive statistics. The motivation for considering THDE in the $f(Q,C)$ gravity framework \cite{AP,AP1} for the current study lies in unifying non-extensive thermodynamic principles with geometric modifications of gravity to effectively describe the late-time accelerated expansion of the Universe. The paper is structured as follows: Section II discusses symmetric teleparallel gravity, including its geometric structure and the impact of non-metricity in changing gravitational field equations. Section III describes the $f(Q,C)$ gravity theory, which extends symmetric teleparallel gravity by including the non-metricity scalar $Q$ and the boundary term $C$, resulting in a more generalized framework for modified gravity models. Section IV elaborates on Tsallis holographic dark energy, presenting the thermodynamic foundations and explaining how the Tsallis entropy framework modifies the traditional holographic dark energy paradigm. Section V reconstructs the $f(Q,C)$ gravity model with Tsallis holographic dark energy using a power-law scale factor, deriving key cosmological quantities to analyze the late time dynamics of the Universe. Section VI presents an observational investigation of Tsallis holographic dark energy inside the $f(Q,C)$ gravity framework, including a thorough discussion on the consistency of the model with observable datasets and implications for cosmic acceleration. Finally, Section VII concludes the paper by summarizing key findings and discussing their implications for cosmology. 

\section{SYMMETRIC TELEPARALLEL GRAVITY}
Symmetric Teleparallel Gravity is a gravitational theory developed within a spacetime geometry with vanishing curvature and torsion, in which gravity is attributed to non-metricity. In this section, we analyze a gravitational model characterized by the four-dimensional metric tensor $g_{\mu\nu}$ and the covariant derivative $\bigtriangledown_{\lambda}$, which is defined using the general connection $\Gamma^{\kappa}_{\mu\nu}$, such that the auto parallels are described as \cite{1}
\begin{equation}
    \frac{d^{2}x^{\mu}}{ds^2}+\Gamma^\mu_{\kappa\nu}\frac{dx^\kappa}{ds}\frac{dx^\nu}{ds}=0.
    \label{1}
\end{equation}
The nature of the geometry is determined by the connection $\Gamma^{\kappa}_{\mu\nu}$. The Riemann tensor can be defined for the general connection as
\begin{equation}
R^\kappa_{\lambda\mu\nu} = \frac{\partial \Gamma^\kappa_{\lambda\nu}}{\partial x^\mu} - \frac{\partial \Gamma^\kappa_{\lambda\mu}}{\partial x^\nu} + \Gamma^\sigma_{\lambda\nu} \Gamma^\kappa_{\mu\sigma} - \Gamma^\sigma_{\lambda\mu} \Gamma^\kappa_{\mu\sigma},
    \label{2}
\end{equation}
the torsion tensor as
\begin{equation}
T^\lambda_{\mu\nu} = \Gamma^\lambda_{\mu\nu} - \Gamma^\lambda_{\nu\mu},
    \label{3}
\end{equation}
and the non-metricity tensor as \cite{2}
\begin{equation}
Q_\lambda{\mu\nu} = \nabla_\lambda g_{\mu\nu}=\frac{\partial g_{\mu\nu}}{\partial x^\lambda} - \Gamma^\sigma_{\lambda\mu} g_{\sigma\nu} - \Gamma^\sigma_{\lambda\nu} g_{\mu\sigma}.
    \label{4}
\end{equation}
The Levi-Civita connection, represented by the symbol $\tilde{\Gamma}\mu\nu^\kappa$, is known as $\Gamma^\kappa_{\mu\nu}$ in General Relativity. Consequently, $T_{\mu\nu^{\lambda}}=0$ and $Q_{\lambda\mu\nu}=0$ in this framework. Thus, the Ricci scalar $R$ is the primary scalar in general relativity.

However, the antisymmetric Weitzenb$"$ock connection takes the place of the connection $\Gamma^\kappa_{\mu\nu}$ in the Teleparallel Equivalent of General Relativity (TEGR) \cite{3}, leading to $R^{\kappa}_{;\lambda\mu\nu}=0$ and $Q_{\lambda\mu\nu}=0$. In this context, the fundamental geometric object in teleparallel gravity is the torsion scalar $T$.

In the STGR theory, $\Gamma^\kappa_{\mu\nu}$ has the characteristics of being torsionless and flat. This means that $T_{\mu\nu^{\lambda}}=0$ and $R^{\kappa}_{;\lambda\mu\nu}=0$. It also inherits the symmetries of the metric tensor $g_{\mu\nu}$. As a result, the non-metricity scalar $Q$ defined as \cite{4}
\begin{equation}
Q=Q_{\lambda\mu\nu}P^{\lambda\mu\nu},
    \label{5}
\end{equation}
is considered the fundamental geometric quantity of gravity.

The tensor $P^{\lambda\mu\nu}$ is defined as \cite{4}
\begin{equation}
P^{\lambda}_{\mu\nu} = -\frac{1}{4} Q^{\lambda}_{\mu\nu} + \frac{1}{2} Q^{~~~\lambda}_{(\mu~\nu)} + \frac{1}{4} (Q^\lambda -\bar{Q}^\lambda )g_{\mu\nu} - \frac{1}{4} \delta^\lambda_{~(\mu}Q_{\nu)},
    \label{6}
\end{equation}
which is written with the help of the traces{\footnote{Symmetrization is indicated by parenthesis in the indices, i.e. $A_{(\mu\nu)}=\frac{1}{2}(A_{\mu\nu}+A_{\nu\mu})$; and $\delta^{\mu}_{~\nu}$ is the Kroncker delta.}} $Q_\mu=Q_{\mu\nu}^{~~~\nu}$ and $\bar{Q}_{}\mu=Q^\nu_{~\mu\nu}$. 

The non-metricity scalar $Q$ for a symmetric and flat connection $\Gamma^\kappa_{\mu\nu}$ differs from the Ricci scalar $R$ for the Levi-Civita connection $\tilde{\Gamma}^\kappa_{\mu\nu}$ of the metric tensor $g_{\mu\nu}$ by a boundary term $C$, which is defined as $C=R-Q$. 

The gravitational Action Integral of STGR is given by \cite{4}
\begin{equation}
\int d^{4}x\sqrt{-g}Q\simeq\int d^{4}x\sqrt{-g}R+ \text{boundary terms}.
    \label{7}
\end{equation}   
Based on this, STGR is dynamically identical to GR. This equivalency is lost, though, when nonlinear components of the non-metricity scalar $Q$ are incorporated into the gravitational action, as in $f(Q)$-gravity. However, the corresponding gravitational theory has no dynamical equivalence to either General Relativity or its generalization, $f(R)$-gravity. 

The Action Integral in symmetric teleparallel $f(Q)$-gravity is defined as\cite{5,jimenez2020born}
\begin{equation}
S_{f(Q)}=\int d^{4}x\sqrt{-g}f(Q).
    \label{8}
\end{equation}
The resulting field equations are as follows:
\begin{equation}
\frac{2}{\sqrt{-g}} \nabla_\lambda \left( \sqrt{-g} f_{,Q} P^\lambda_{\ \mu\nu} \right) - \frac{1}{2} f(Q) g_{\mu\nu} + f_{,Q} (P_{\mu\rho\sigma} Q^{\rho\sigma}_{\ \nu} - 2 Q_{\rho \sigma\mu} P^{\rho\sigma}_{\nu}) = 0
    \label{9}
\end{equation}
\begin{equation}
\nabla_\mu \nabla_\nu \left( \sqrt{-g} f_{,Q} P^{\mu\nu}_{~~\sigma} \right) = 0
    \label{10}
\end{equation}
The modified Einstein field equations in $f(Q)$-gravity are represented by Eq.(\ref{9}), and the equation of motion for the connection is defined by Eq.(\ref{10}). If Eq.(\ref{10}) is always true for a specific connection, that connection is called the ``coincidence gauge.'' If Eq.(\ref{10}) is not always true for a particular connection, that connection is defined within the so-called ``non-coincidence gauge,'' as explained in \cite{jimenez2020born}.

\section{$f(Q,C)$ gravity}

{In this Section, we present a brief overview of $f(Q,C)$ \cite{de2024non,SS,SS1} gravity, as in the subsequent Sections, we are going to present a rigorous reconstruction cosmology as already stated. We have already elaborated in the Introduction how in recent years, alternative theories of gravity have gained significant interest to  study cosmic acceleration and dark energy without invoking unknown matter components. Among the various candidates of modified gravity, symmetric teleparallel gravity, as explained in the previous Section, is one kind of gravity, where gravity is attributed to non-metricity rather than curvature or torsion. This has emerged as a promising framework \cite{4, beltran2019geometrical}. In this context, the non-metricity scalar \( Q \) plays a central role in characterizing the gravitational interaction. A generalization of this theory is known as \( f(Q) \) gravity, where the action is constructed as an arbitrary function of \( Q \) \cite{jimenez2018teleparallel}. Further extending this idea, the \( f(Q,C) \) gravity theory incorporates both the non-metricity scalar \( Q \) and a boundary term \( C \), analogous to the boundary term in \( f(R) \) theories of curvature-based gravity \cite{Shabani, Samaddar}. The incorporation of \( C \) gives the scope for a continuous interpolation between different formulations of gravity and provides a more sophisticated dynamics by potentially addressing both the phenomena of early and late-time cosmology. This generalized framework paves avenues for studying modified gravitational interactions and exploring deviations from general relativity.} A boundary term is incorporated into the gravitational action integral in a recent extension of the $f(Q,C)$ theory\cite{capozziello2023,de2024non,9}. The gravitational action integral is included in this generalization in the manner described below:
\begin{equation}
S=\int d^{4}x\sqrt{-g}\biggl[\frac{1}{2\kappa}f(Q,C)+\mathcal{L}_{m}\biggr],
    \label{11}
\end{equation}
where $g$ is the determinant of the metric tensor $g_{\mu\nu}$, $\kappa=8\pi G_{N}=1$, and $G_{N}$ denotes the gravitational constant. 

To build a realistic cosmological model, we consider a matter action $S_{m}$, connected to the energy-momentum tensor $\Theta_{\mu\nu}$. The following Friedmann equations result from varying the total action $S$, as \cite{de2024non} demonstrated:
\begin{equation}
\begin{array}{c}
\kappa T_{\mu\nu}=-\frac{f}{2}g_{\mu\nu}+\frac{2}{\sqrt{-g}}\partial_{\eta}(\sqrt{-g}f_{Q}P^{\eta}_{~\mu\nu})+(P_{\mu\alpha\beta}Q_{\nu}^{~\alpha\beta}-2P_{\alpha\beta\nu}Q^{\alpha\beta}_{~~\mu})f_{Q}+\\
\biggl(\frac{C}{2}g_{\mu\nu}-\overset{\circ}{\nabla}_{\mu}\overset{\circ}{\nabla}_{\nu}+g_{\mu\nu}\overset{\circ}{\nabla}^{\alpha}\overset{\circ}{\nabla_\alpha}-2P^\eta_{~\mu\nu}\partial_\eta\biggr)f_C.
\end{array}
    \label{12}
\end{equation}
Eq.(\ref{12}) can be expressed in a covariant manner as:
\begin{equation}
\begin{array}{c}
\kappa T_{\mu\nu}=-\frac{f}{2}g_{\mu\nu}+2P^\eta_{~\mu\nu}\overset{\circ}{\nabla}_{\eta}(f_{Q}-f_{C})+\biggl(\overset{\circ}{G}_{\mu\nu}+\frac{Q}{2}g_{\mu\nu}\biggr)f_{Q}+\\
\biggl(\frac{C}{2}g_{\mu\nu}-\overset{\circ}{\nabla}_{\mu}\overset{\circ}{\nabla}_{\nu}+g_{\mu\nu}\overset{\circ}{\nabla}^{\alpha}\overset{\circ}{\nabla_\alpha}\biggr)f_C.
\end{array}
    \label{13}
\end{equation}
The effective stress-energy tensor is defined as follows:
\begin{equation}
\begin{array}{c}
T_{\mu\nu}^{eff}=T_{\mu\nu}+\frac{1}{\kappa}\biggl[\frac{f}{2}g_{\mu\nu}-2P^\eta_{~\mu\nu}\overset{\circ}{\nabla}_{\eta}(f_{Q}-f_{C})-\frac{Qf_Q}{2}g_{\mu\nu}-\\
\biggl(\frac{C}{2}g_{\mu\nu}-\overset{\circ}{\nabla}_{\mu}\overset{\circ}{\nabla}_{\nu}+g_{\mu\nu}\overset{\circ}{\nabla}^{\alpha}\overset{\circ}{\nabla_\alpha}\biggr)f_C\biggr].
\end{array}
    \label{14}
\end{equation}
To create an equation that resembles that of GR,
\begin{equation}
\overset{\circ}{G}_{\mu\nu}=\frac{\kappa}{f_Q}T_{\mu\nu}^{eff}.
    \label{15}
\end{equation}
This section introduces $f(Q,C)$ cosmology and examines the application of $f(Q,C)$ gravity in a cosmological framework. Our analysis examines a homogeneous and isotropic flat Friedmann–Lemaître–Robertson–Walker (FLRW) spacetime,  described by the line element in Cartesian coordinates
\begin{equation}
ds^2=-dt^2+a^2(t)[dx^{2}+dy^2+dz^2],
\label{16}
\end{equation}
where the scale factor is denoted by $a(t)$ and is connected to the Hubble parameter $H$ via its first-time derivative, $H(t)=\frac{\dot{a}(t)}{a(t)}$.

Following this section, it has been demonstrated that within the framework of $f(Q,C)$ gravity \cite{SS,SS1}, an additional effective sector of geometrical origin can be derived, as shown in Eq.(\ref{14}). When examined in a cosmological context, this term can be interpreted as an effective DE sector that possesses an energy-momentum tensor.
\begin{equation}
    \begin{array}{cc}
    T_{\mu\nu}^{DE}=\frac{1}{f_{Q}}\biggl[\frac{f}{2}g_{\mu\nu}-2P^\eta_{~\mu\nu}\overset{\circ}{\nabla}_{\eta}(f_{Q}-f_{C})-\frac{Qf_Q}{2}g_{\mu\nu}-\\
\biggl(\frac{C}{2}g_{\mu\nu}-\overset{\circ}{\nabla}_{\mu}\overset{\circ}{\nabla}_{\nu}+g_{\mu\nu}\overset{\circ}{\nabla}^{\alpha}\overset{\circ}{\nabla_\alpha}\biggr)f_C\biggr],
    \end{array}
    \label{17}
\end{equation}
\begin{equation}
\overset{\circ}{R}=6(2H^{2}+\dot{H}),~Q=-6H^{2},~C=6(3H^{2}+\dot{H}).
\label{18}
\end{equation}
We consider a vanishing affine connection $(\Gamma^{\eta}_{\mu\nu}=0)$ while fixing the coincident gauge. From this data, we can derive our Friedmann-like equations as follows:
\begin{equation}
3H^{2}=\kappa(\rho_{m}+\rho_{r}+\rho_{DE}),
    \label{19}
\end{equation}
\begin{equation}
-2\dot{H}-3H^{2}=\kappa\biggl(\frac{\rho_{r}}{3}+p_{DE}\biggr),
    \label{20}
\end{equation}
where $\rho_{m}$, $\rho_{r}$, $\rho_{DE}$, and $p_{DE}$ represent the densities of matter, radiation, DE, and the pressure of DE, respectively, treated as a perfect fluid. In the absence of any interactions between non-relativistic matter and radiation, each component follows its own conservation laws. These are expressed as: \\
For matter: 
\begin{equation}
\dot{\rho}_{m} + 3 H \rho_{m} = 0,
\label{020}
\end{equation}
and for radiation: 
\begin{equation}
\dot{\rho}_{r} + 4 H \rho_{r} = 0.
\label{0020}
\end{equation}
However, we have considered the contribution of dark energy and dark matter at late time. Additionally, we have defined the effective density and pressure for DE as follows:
\begin{equation}
\rho_{DE}=\frac{1}{\kappa}[3 H^{2}(1-2f_{Q})-\frac{f}{2}+(9 H^{2}+3\dot{H})f_{C}-3H\dot{f}_{C}],
    \label{21}
\end{equation}
\begin{equation}
    p_{DE}=\frac{1}{\kappa}[-2 \dot{H}(1-f_{Q})-3 H^{2}(1-2f_{Q})+\frac{f}{2}+2H\dot{f}_{Q}-(9H^{2}+3\dot{H})f_{C}+\ddot{f}_{C}]
    \label{22}
\end{equation}
Since standard matter is conserved independently, it can be deduced from Eqs. (\ref{21}) and (\ref{22}) that the DE density and pressure follow the standard evolution equation
\begin{equation}
\dot{\rho}_{DE}+3H(\rho_{DE}+p_{DE})=0.
    \label{23}
\end{equation}
The equation of state (EoS) parameter can be defined for the DE as
\begin{equation}
    w_{DE}=\frac{p_{DE}}{\rho_{DE}}
    \label{24}.
\end{equation}
In summary, the $f(Q,C)$ gravity framework provides a geometrically motivated modification to General Relativity, forming a robust foundation for embedding generalized dark energy models.  As per the theme of the current study, we intend to integrate Tsallis Holographic Dark Energy (THDE) with it. We discuss THDE in the next section.

\section{Tsallis holographic dark energy}
The definition and derivation of the standard HDE density $(\rho_{DE}=3c^{2}m_{p}^{2}L^{-2})$ are based on the entropy-area relationship $S\sim A\sim L^{2}$ of black holes, with $A=4\pi L^{2}$ representing the area of the horizon \cite{T1}. However, considering quantum factors \cite{T2,T3}, this definition of HDE can be modified. Tsallis and Cirto \cite{T4} demonstrated that the horizon entropy of a black hole may be altered as
\begin{equation}
S_{\delta}=\gamma A^{\delta},
    \label{25}
\end{equation}
where $\delta$ and $\gamma$ are nonadditivity parameter and an unknown constant, respectively \cite{T4}. The Bekenstein entropy can be obtained in the appropriate limits where $\delta = 1$ and $\gamma = \frac{1}{4G}$ (using units where $h = \kappa_{B} = c = 1$). In this scenario, the system can be described by a standard probability distribution, and the power-law probability distribution becomes irrelevant. This connection is further supported by findings in quantum gravity \cite{T5} and leads to intriguing results in holographic and cosmic configurations \cite{T6,T7,T8,T9,T10}. According to the holographic principle, an infrared cutoff should constrain a physical system's degrees of freedom. Furthermore, these degrees of freedom should scale with the system's enclosing area rather than its volume \cite{tHooft1993dimensional,T12}. Cohen et al. \cite{T1} proposed a relationship between the system's entropy $S$ and the ultraviolet $\Lambda$ and infrared $L$ cutoffs
\begin{equation}
L^{3}\Lambda^{3}\leq(S)^{\frac{3}{4}}.
    \label{26}
\end{equation}
On combining Eq.(\ref{25}) with Eq.(\ref{26}), we have \cite{T1}
\begin{equation}
\Lambda^{4}\leq\gamma(4\pi)^{\delta}L^{2\delta-4},
    \label{27}
\end{equation}
where $\Lambda^{4}$ represents the vacuum energy density, the energy density of DE $(\rho_{DE})$ in the HDE hypothesis \cite{T14,T15,T16}. The THDE is proposed based on the aforementioned inequality as
\begin{equation}
\rho_{D}=\zeta L^{2\delta-4},
    \label{28}
\end{equation}
where $\zeta$ is a parameter \cite{T14,T15,T16}. In the subsequent sections, we would explore this formulation for the study of further cosmology under the purview of $f(Q,C)$ gravity. 

\section{THDE in $f(Q,C)$ gravity}
This section is aimed at exploring the cosmology of THDE in $f(Q,C)$ gravity framework. Before going into further details, let us discuss the motivation behind considering THDE in $f(Q,C)$ framework. To study the same, we choose the scale factor $a(t)$ in the power-law form. The THDE model, motivated by non-extensive entropy concerns, presents a generalized framework capable of predicting dynamical dark energy behavior beyond the cosmological constant \cite{tavayef2018tsallis, saridakis2018tsallis1}. Simultaneously, modified gravity theories based on non-metricity, such as $f(Q)$ and its expansions to $f(Q,C)$ gravity, have emerged as feasible frameworks to explain the rapid expansion without invoking exotic matter components \cite{Bhagat2025,Harko2018}. The $f(Q,C)$ extension enables rigorous cosmic dynamics by including the boundary term $C$ alongside the non-metricity scalar $Q$, resulting in alterations similar to those seen in $f(R)$ gravity. The THDE technique is combined with the $f(Q,C)$ gravitational framework to solve late-time acceleration while being consistent with thermodynamic principles and gravity's geometric alterations. Previous studies integrating entropy-based dark energy models into modified gravity frameworks further motivate this approach \cite{Sultana2024,Karami2011}.

{At this juncture let us have a note on the motivation behind considering THDE in $f(Q,C)$ framework with scale factor in the form  \( a(t) = a_0 t^n \). It has already been elaborated that studying THDE in the context of \( f(Q,C) \) gravity is a compelling approach to combine entropy-based dark energy models and geometrically motivated modified gravity, two avenues to explain late-time acceleration. To understand the time evolution of cosmological parameters, a power-law scale factor of the form \( a(t) = a_0 t^n \) is often used in cosmological modelling. This ansatz not only simplifies the field equations but also effectively models different phases of cosmic expansion, including the radiation-dominated, matter-dominated, and accelerated eras depending on the value of \( n \) \cite{nojiri2011unified}.} This power-law form of scale factor leads to the Hubble parameter, non-metricity scalar and boundary term as
\begin{equation}
H=\frac{n}{t},
    \label{30}
\end{equation}
\begin{equation}
Q=-\frac{6n^2}{t^2},
    \label{31}
\end{equation}
\begin{equation}
C=\frac{6n(3n-1)}{t^2},
    \label{32}
\end{equation}
respectively. The first derivative of $H$, $Q$ and $C$ are:
\begin{equation}
\dot{H}=-\frac{n}{t^{2}},
    \label{33}
\end{equation}
\begin{equation}
\dot{Q}=\frac{12n^2}{t^{3}},
    \label{34}
\end{equation}
\begin{equation}
\dot{C}=-\frac{12n(3n-1)}{t^3}.
    \label{35}
\end{equation}
Hence we have the expressions for $f_{Q}$, $\dot{f}_Q$, $f_C$, $\dot{f}_C$ and $\ddot{f}_C$ in terms of $\dot{f}$, $\ddot{f}$ and $\dddot{f}$ as:
\begin{equation}
f_Q=\frac{\dot{f}t^{3}}{12n^2},
    \label{36}
\end{equation}
\begin{equation}
\dot{f}_Q=\frac{1}{12 n^2}\left(t^3\ddot{f} +3 \dot{f} t^2\right),
    \label{37}
\end{equation}
\begin{equation}
f_C=-\frac{\dot{f}t^{3}}{12n(3n-1)},
    \label{38}
\end{equation}
\begin{equation}
\dot{f}_C=-\frac{1}{12 n(3n-1)}\left(t^3 \ddot{f}+3 \dot{f} t^2\right),
    \label{39}
\end{equation}
\begin{equation}
\ddot{f}_C=-\frac{1}{12 n(3n-1)}\left(t^3 \dddot{f}+3\ddot{f}t^2+6 t \dot{f}+3 t^2 \ddot{f}\right).
\label{40}
\end{equation}
By substituting Eq.(\ref{28}) to the left-hand side of Eq.(\ref{21}) and Eqs. (\ref{30}), (\ref{33}), (\ref{36}), (\ref{38}), (\ref{39}) to the right-hand side of  Eq.(\ref{21}), we have the reconstructed $f$ as
\begin{equation}
\begin{split}
f_{Q,C}(t) = & \, \frac{9n^3}{(1-3n)t^2} + \frac{3n^2}{(-1+3n)t^2} 
+ \frac{2\zeta n^4 \left(\frac{n}{t}\right)^{-2\delta}}{t^4 \left(1+(3-2\delta)\delta+3n(-5+3\delta)\right)} \\
& - \frac{6\zeta n^5 \left(\frac{n}{t}\right)^{-2\delta}}{t^4 \left(1+(3-2\delta)\delta+3n(-5+3\delta)\right)} \\
& + t^{\frac{1}{2}\left(-5 - \sqrt{13+\frac{4}{1-3n} - 27n} \sqrt{1-3n} + 9n\right)}
\left( C_1 + t^{\sqrt{13+\frac{4}{1-3n} - 27n} \sqrt{1-3n}} C_2 \right).
\end{split}
\label{41}
\end{equation}
\textbf{Some limiting cases}: In the above equation, where $C_1$ and $C_2$ are constants of integration, and the reconstructed $f(Q,C)$ appears as a function of cosmic time $t$, since both $Q$ and $C$ have been re-expressed in terms of $t$. Therefore, we write the function as $f(t)$. The reconstructed form of $f(t)$ reflects the dynamic behavior available through different higher order derivatives of $f(t)$ namely $f_{Q}$, $\dot{f}_Q$, $f_C$, $\dot{f}_C$ and $\ddot{f}_C$ and this reconstructed modified gravity serves as a key ingredient in further analysis of the dynamics of the model. Let us have some insight into the limiting conditions of Eq.(\ref{41}). If $n\to 0$ (i.e. $n\ll 1$), we observe that the first 4 terms vanish and $f(t)$ is dominated by the power law term involving $C_{1}$ and $C_{2}$. Hence, in this limiting condition 
\begin{equation}
f_{Q,C}(t)=t^{\frac{1}{2}(-5-\sqrt{17})}(C_1+t^{\sqrt{17}}C_2).
    \label{041}
\end{equation}
We consider another limiting scenario pertaining to the very late stage, for which let us consider $t\to \infty$. In this case, the first four terms vanish because $2\delta-4 <0$. Now the last term depends on the sign of the exponent of $t$. If the exponent is negative, the term will tend to zero with the increase in $t$. On the contrary, in case of the positive exponent, $f(t)$ will grow with the increase in $t$. Considering the above two limiting cases, we understand that the precise behavior of $f(t)$ significantly depends on the values of $n$. As we have considered the limiting case of the very late stage, we consider the very early stage of $f(t)$, i.e. $t\to 0$. In this scenario, the dominant contribution to $f(t)$ will be due to the terms involving $t^{-2}$ and $t^{-4}$. However, the contribution due to the last term, where $t$ has exponents, will depend on the choice of $n$. It may be noted that we have chosen $\delta$ to be a small positive quantity. Consolidating the above outcomes, we get the significance of the values of $n$ on the precise behavior of reconstructed $f(Q,C)$. We can re-express Eq.(\ref{041}) as a function of non-metricity scalar $Q$ and boundary term $C$ as 
\begin{equation}
f(Q,C)=\frac{6^{\frac{1}{4}\left(-5-\sqrt{17}\right)} \left(-\frac{Q}{(C+3 Q)^2}\right)^{\frac{1}{4} \left(3-\sqrt{17}\right)} (C+3 Q)^4 \left(C_1+\left(-\frac{6Q}{(C+3 Q)^2}\right)^{\frac{\sqrt{17}}{2}} C_2\right)}{Q^2}.
    \label{f41}
\end{equation}

From the reconstructed $f(Q,C)$ (Eq.(\ref{41})), its time derivatives, and the equations mentioned above, we have reconstructed density and pressure for THDE in $f(Q,C)$ gravity as
\begin{equation}
\begin{array}{cc}
  \rho_{THDE}=\frac{\zeta n^4 \left(\frac{n}{t}\right)^{-2 \delta }}{t^4},
\end{array}
    \label{42}
\end{equation}
\begin{equation}
\begin{array}{cc}
  p_{THDE}=\left(\left(\frac{n}{t}\right)^{-2 \delta } t^{\frac{1}{2} \left(-9-\sqrt{13+\frac{4}{1-3 n}-27 n} \sqrt{1-3 n}\right)}\left(-C_{1} \left(5+\sqrt{13+\frac{4}{1-3
n}-27 n} \sqrt{1-3 n}-9 n\right)\right.\right.\\
 \left(\frac{n}{t}\right)^{2 \delta } t^{2+\frac{9 n}{2}} (1+(3-2 \delta ) \delta +3 n (-5+3 \delta ))+t^{\frac{1}{2} \left(1+\sqrt{13+\frac{4}{1-3
n}-27 n} \sqrt{1-3 n}\right)} \\
\left(\left(\frac{n}{t}\right)^{2 \delta } t \left(12 n^2 t+C_{2} \left(-5+\sqrt{13+\frac{4}{1-3 n}-27 n} \sqrt{1-3 n}\right) t^{\frac{1}{2}
\left(1+\sqrt{13+\frac{4}{1-3 n}-27 n} \sqrt{1-3 n}+9 n\right)}+\right.\right.\\
\left.9 C_{2} n t^{\frac{1}{2} \left(1+\sqrt{13+\frac{4}{1-3 n}-27 n} \sqrt{1-3 n}+9 n\right)}\right) (1+(3-2 \delta ) \delta +3 n (-5+3
\delta ))-\\
\left.\left.\left.\left.4 \zeta n^4 \left(-2 (-2+\delta ) \delta  (-3+2 \delta )+9 n^2 (-5+3 \delta )+3 n (17+\delta  (-17+4 \delta ))\right)\right)\right)\right)\right.\\
(12 n (1+(3-2 \delta ) \delta +3 n (-5+3 \delta )))^{-1}.
\end{array}
    \label{43}
\end{equation}
respectively. From Eq.(\ref{020}), we have the reconstructed density for dark matter as
\begin{equation}
    \rho_{m}=t^{-3n}C_3.
    \label{45}
\end{equation}
By implementing Eqs. (\ref{42}), (\ref{43}) and (\ref{45}) in the Friedmann equations, and by taking the relation $t=\left(\frac{1}{a_0(1+z)}\right)^{\frac{1}{n}}$ we have the reconstructed Hubble parameter for THDE in $f(Q,C)$ gravity in terms of redshift $z$ as
\begin{equation}
\begin{array}{cc}
 H(z)=\frac{1}{24} \left(\frac{1}{a_{0}(1+z)}\right)^{-3/n} \\
\left(\frac{C_1 \left(1+\sqrt{13+\frac{4}{1-3 n}-27 n} \sqrt{1-3 n}+3 n\right) \left(\left(\frac{1}{a_{0}(1+z)}\right)^{\frac{1}{n}}\right)^{\frac{1}{2}
\left(3-\sqrt{13+\frac{4}{1-3 n}-27 n} \sqrt{1-3 n}+9 n\right)}}{n (-2+3 n)}-\right.\\
\frac{C_2 \left(-1+\sqrt{13+\frac{4}{1-3 n}-27 n} \sqrt{1-3 n}-3 n\right) \left(\left(\frac{1}{a_{0}(1+z)}\right)^{\frac{1}{n}}\right)^{\frac{1}{2}
\left(3+\sqrt{13+\frac{4}{1-3 n}-27 n} \sqrt{1-3 n}+9 n\right)}}{n (-2+3 n)}+\\
12 n \left(\frac{1}{a_{0}(1+z)}\right)^{2/n}+\frac{12 C_3 \left(\left(\frac{1}{a_{0}(1+z)}\right)^{\frac{1}{n}}\right)^{-3
n} \left(\frac{1}{a_{0}(1+z)}\right)^{4/n}}{-1+3 n}+\\
\left.\frac{8 \zeta n^3 \left(n \left(\frac{1}{a_{0}(1+z)}\right)^{-1/n}\right)^{-2 \delta } (-2+\delta ) ((3-2 \delta ) \delta +3 n
(-4+3 \delta ))}{(-3+2 \delta ) (1+(3-2 \delta ) \delta +3 n (-5+3 \delta ))}\right)+C_4.
\end{array}
    \label{46}
\end{equation}
The expression for reconstructed Hubble parameter $H$ as presented in Eq.(\ref{46}) gets the term involving $C_1$ and $C_2$ from the reconstruction of $f(Q,C)$, and therefore, they incorporate the effects of initial conditions. The term $C_3$ arises from the consideration of dark matter. It should be noted that there are terms to the power of $\delta$ reflecting the influence of this parameter arising from Tsallis entropy modification. This indicates the deviation of the evolutionary pattern of $H$ for THDE under $f(Q,C)$ framework from standard Hubble evolution at early and late Universe. The additive constant $C_4$ represents an integration constant that could be associated with effective vacuum energy density. Overall, the complete expression of $H$ gives us the possibility of studying the Hubble dynamics with a set of free parameters, and hence our parameter space is expanded to $\Theta=(n,\delta,\zeta,a_0,C_1,C_2,C_3,C_4)$.

{Let us mention here the physical dimensions of the constants of integration appearing in the above equations. Since $H(z)$ represents the Hubble parameter, which has the dimension inverse of cosmic time $t$. The constant $C_4$ appears in Eq.(\ref{46}) as an additive contribution and therefore has the same dimension as $H_0$. On the other hand, the constants $C_1$, $C_2$, and $C_3$ arise from the integration of the reconstructed gravitational function and therefore can be thought of controlling the magnitude of geometric corrections. In numerical methods, such constraints are effectively treated in a normalized form with respect to the Hubble Scale, so that their contributions represent dimensionless corrections to the evolution of the universe.}

\begin{figure}
\begin{center}
\includegraphics[height=1.5in]{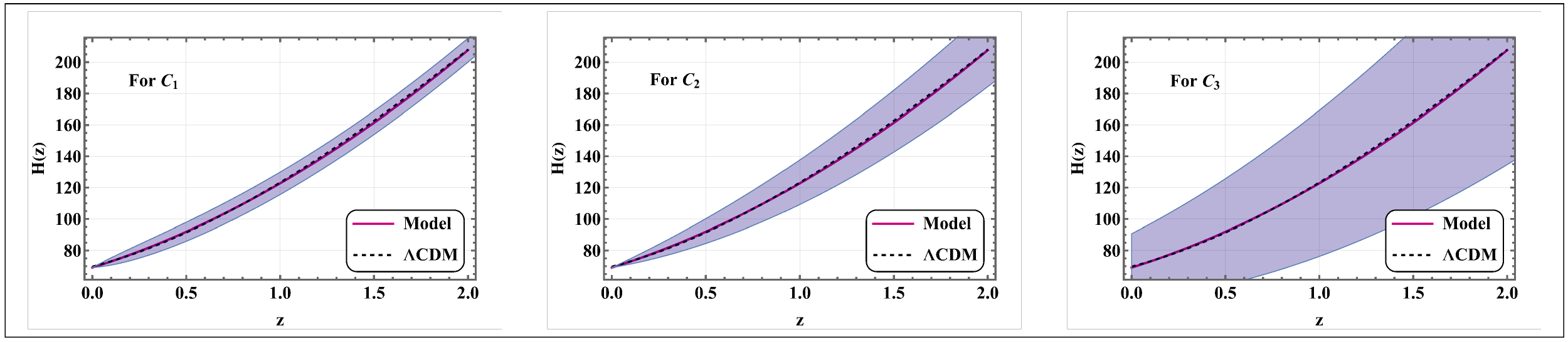}
\caption{{Theoretical plots showing how different values of the constants $C_1,C_2,C_3$ affect the evolution of $H(z)$.}}
\label{Diff}
\end{center}
\end{figure}

{In Figure~\ref{Diff}, we have plotted the evolutionary behavior of $H(z)$ for the reconstructed THDE in the framework of $f(Q,C)$ gravity for different choices of the constant of integration. Each panel corresponds to a distinct value of the constant, and accordingly, theoretical plots of $H(z)$ are generated to view how these parameters influence the cosmic expansion history. The solid curve represents the prediction from the $f(Q,C)$-THDE model under consideration. The dashed curve pertains to the standard $\Lambda$CDM scenario. The shaded region represents the uncertainty band, and it arises from the parameter space associated with the model. The theoretical curves presented in Figure~\ref{Diff} show smooth monotonic increasing behavior with $z$, implying the decrease of $H(z)$ with the evolution of the universe. For the different choices of the constant of integration, the basic behavior of the curves is the same, but the uncertainty band widens from $C_1$ to $C_3$. This indicates the sensitivity of the model to the choice of constant of integration. However, the close overlap between the model curve and $\Lambda$CDM over the full redshift range indicates that the reconstructed model remains observationally viable. The constants $C_1$ and $C_2$ are coming from reconstructed $f(Q,C)$ as reflected in Eq.(\ref{41}). Thus, the widening of the uncertainty band indicates that the constants of integration modify the strength of geometric corrections arising from the $f(Q,C)$ sector.We would like to mention that for the constant $C_4$, we have utilized Eq.(\ref{46}) by choosing $z=0$, giving $H_0$, the present-day value of Hubble parameter, and accordingly taking $C_4=H_0-\nu_0$ where $\nu_0$ represents the first term of the RHS of Eq.(\ref{46}) for $z=0$. Furthermore, while generating Figure~\ref{Diff}, we analyzed the sensitivity to $C_1$ (Left panel), considering the range $-2\times 10^5 < C_1 < 2\times 10^5$. The middle panel, which depicts sensitivity to $C_2$, has a theoretical range of $-20 < C_2 < 20$. The right panel illustrates $C_3$- sensitivity with range $-2< C_3 < 2$. In view of the choice of $C_1$, we can say that despite choosing a significantly large range of $C_1$, the sigma band is notably narrow, indicating the low sensitivity of the theoretical model to the constant $C_1$. On the contrary, for $C_2$, a much smaller range is chosen. The width of the band is comparatively higher. This indicates greater sensitivity to $C_2$ than $C_1$. Finally, the right panel shows a higher sensitivity to $C_3$, where a significantly large width of the band is available for a very narrow range of $C_3$.}

\section{Observational Analysis of Tsallis Holographic Dark Energy in $f(Q,C)$ gravity}

This section is dedicated to the observational viability of the THDE model within the framework of \( f(Q,C) \) gravity. The Markov Chain Monte Carlo (MCMC) method would be employed to efficiently explore the parameter space and obtain the best-fit values along with their confidence intervals. 

\subsection{Markov Chain Monte Carlo (MCMC) analysis of the model with Observational Data}
In this subsection, we perform a comprehensive parameter estimation for a reconstructed cosmological model inspired by THDE in the context of modified $f(Q,C)$ gravity, using an MCMC approach. In this context, let us mention that although the integration constants $C_1$, $C_2$, $C_3$, and $C_4$ were earlier mentioned within the parameter space $\Theta$, those have not been constrained here; instead, they have been utilized in the following way. By choosing $C_1$ and $C_2$ to be unity, we have given equal preference to the different power law modes by not favoring one over the other, and treating them as symmetrically contributing to the expansion. By setting $C_3$ equal to zero, we have allowed the corresponding term to be dominant in the late time acceleration of the Universe. Finally, boundary condition $H_0$ (Hubble constant) is set by eliminating $C_4$ which makes our reconstructed model directly comparable to $\Lambda$CDM and ensures physical realism at $z=0$. Hence, the model is governed by five principal parameters: $H_0$, $a_0$, $n$, $\delta$, and $\zeta$ that appear in the correction term characterizing deviations from standard cosmology. The Hubble function $H(z)$ for our model is reconstructed from the modified background dynamics of the geometric theory. {\color{black} The simplified analytical form of $H(z)$ derived from Eq.(\ref{46}) is given by
\begin{align}
H(z)=H_0+\frac{1}{24}\Bigg\{&
\frac{8\zeta(\delta-2)n^{3-2\delta}
\big[\delta(3-2\delta)+3(3\delta-4)n\big]}
{(2\delta-3)\big[\delta(3-2\delta)+3(3\delta-5)n+1\big]}
\,a_0^{\frac{3-2\delta}{n}}
\Big[(1+z)^{\frac{3-2\delta}{n}}-1\Big]
\nonumber\\[6pt]
&+12n\,a_0^{\frac{1}{n}}
\Big[(1+z)^{\frac{1}{n}}-1\Big]
\nonumber\\[6pt]
&+\frac{3n-\Gamma+1}{n(3n-2)}\,
a_0^{\frac{3-\Gamma}{2n}}
\Big[(1+z)^{\frac{3-\Gamma}{2n}}-1\Big]
\nonumber\\[6pt]
&+\frac{3n+\Gamma+1}{n(3n-2)}\,
a_0^{\frac{3+\Gamma}{2n}}
\Big[(1+z)^{\frac{3+\Gamma}{2n}}-1\Big]
\Bigg\},
\end{align}
\begin{equation}
\Gamma=\sqrt{1-3n}\,
\sqrt{-27n+\frac{4}{1-3n}+13}.
\end{equation}
}
The parameters $a_0$, $n$, $\delta$, and $\zeta$ play an important role in shaping the evolution of $H(z)$, thereby influencing the model compatibility with observational data. For benchmarking purposes, we also consider the standard $\Lambda$CDM model, described by the well-known form: $H(z) = H_0 \sqrt{\Omega_m(1+z)^3 + (1 - \Omega_m)}$, with a fixed matter density $\Omega_m = 0.3$.

In order to constrain the cosmological parameter vector 
$\boldsymbol{\Theta} = (H_0, a_0, n, \delta, \zeta)$, 
we employ a Bayesian inference approach based on Markov Chain Monte Carlo (MCMC) techniques 
\cite{mcmc_cosmology,foreman2013emcee,karamanis2021zeus}. This framework enables an efficient exploration of the multidimensional parameter space 
and provides statistically robust estimates of the posterior distributions of the model parameters. Within the Bayesian formalism, the posterior probability distribution is obtained using Bayes’ theorem,
\begin{equation}\label{O1}
P(\boldsymbol{\theta} \mid D, I) =
\frac{P(D \mid \boldsymbol{\theta}, I)\,P(\boldsymbol{\theta} \mid I)}
     {P(D \mid I)},
\end{equation}
where $P(\boldsymbol{\theta} \mid I)$ denotes the prior probability distribution encoding our initial assumptions about the parameters, $P(D \mid \boldsymbol{\theta}, I)$ is the likelihood function corresponding to the observational dataset $D$, and $P(D \mid I)$ represents the Bayesian evidence, which acts as a normalization factor.

The likelihood is constructed from the chi-square statistic associated with the model predictions and the observational data, and is defined as
\begin{equation}\label{O2}
\mathcal{L}(\boldsymbol{\theta}) = \exp\!\left[-\frac{1}{2}\chi^2(\boldsymbol{\theta})\right].
\end{equation}
In this work, we have used an updated set of Hubble parameter measurements, comprising 32 points from Cosmic Chronometers (CC) \cite{O3,CC2,CC3,CC4} obtained via the differential age (DA) method and DESI DR2 Baryon Acoustic Oscillation (BAO) data \cite{karim2025desi,Lodha2025DESI,bhagat2026logarithmic} observations and related techniques. Additionally, we include the Pantheon compilation of 1701 Type Ia supernovae \cite{brout2022pantheon,brout2022pantheonplus,scolnic2022pantheonplus}, which offers precise distance modulus $\mu(z)$ measurements. This combination provides a robust coverage of the expansion history across a wide redshift range. Through this analysis, we aim to test the viability of the THDE model in \( f(Q,C) \) gravity and to assess its consistency with the current observational evidence. The resulting best-fit values and confidence intervals are expected to help in having a critical insight into the cosmological dynamics under the modified gravity scenario considered.

\textbf{Cosmic Chronometers (CC) Dataset:}~The cosmic chronometers method~\cite{O3,CC2,CC3,CC4} offers direct and model-independent determinations of the Hubble expansion rate, $H(z)$, by exploiting the differential aging of massive, passively evolving galaxies. This technique is based on the relation $H(z) = -\frac{1}{1+z}\frac{dz}{dt}$, where the quantity $dz/dt$ is inferred from the relative age evolution of early-type galaxies observed at different redshifts.

In the present analysis, we employ a compilation of 32 CC measurements~\cite{O3,CC2,CC3,CC4,Bhagat2026}, spanning the redshift interval $0.07 \leq z \leq 1.965$, collected from a variety of spectroscopic surveys. The contribution of the CC dataset to the total likelihood is quantified through the chi-square estimator
\begin{equation}\label{O4}
\chi^2_{\rm CC}(\boldsymbol{\theta}) =
\sum_{i=1}^{32}
\frac{\left[H_{\rm th}(z_i;\boldsymbol{\theta}) -
H_{\rm obs}(z_i)\right]^2}
{\sigma_H^2(z_i)},
\end{equation}
where $H_{\rm th}(z_i;\boldsymbol{\theta})$ denotes the theoretical prediction of the Hubble parameter at redshift $z_i$ for a given set of model parameters $\boldsymbol{\theta}$, $H_{\rm obs}(z_i)$ represents the corresponding observational value, and $\sigma_H(z_i)$ is the associated standard deviation.

{\textbf{Pantheon$^{+}$ Supernovae Dataset:} The Pantheon$^{+}$ compilation~\cite{brout2022pantheon,scolnic2022pantheonplus,brout2022pantheonplus} represents a substantial extension of the original Pantheon sample~\cite{SN}, providing one of the most comprehensive collections of Type Ia supernovae (SNe~Ia) observations to date. It includes 1701 light-curve measurements corresponding to 1550 spectroscopically confirmed SNe~Ia. This dataset incorporates improved photometric calibration along with an extended redshift coverage. The sample spans the range $0.00122 \leq z \leq 2.2613$, combining data from both ground-based surveys and space-based missions, including observations from the Hubble Space Telescope.}
{In SNe~Ia cosmology, the primary observable is the distance modulus, which for a given cosmological model with parameters $\boldsymbol{\theta}$ is expressed as
\begin{equation}
\mu_{\rm th}(z;\boldsymbol{\theta}) = 5\log_{10}\!\left[d_L(z;\boldsymbol{\theta})\right] + \mu_0,
\label{O5}
\end{equation}
where $\mu_0$ is a nuisance parameter that encapsulates the absolute magnitude of supernovae and the present value of the Hubble constant. The luminosity distance is defined as
\begin{equation}\label{O6}
d_L(z) = (1+z)\int_{0}^{z}\frac{dz'}{E(z')},
\end{equation}
where $E(z)=H(z)/H_0$ denotes the normalized Hubble expansion rate.}
{To constrain the cosmological parameters, we adopt the standard likelihood formalism commonly used in SNe~Ia analyses. Following Conley et al.~\cite{Conley}, the $\chi^2$ function is constructed as
\begin{equation}
\chi^2 = \Delta D^{T} C^{-1}_{\mathrm{SN}} \Delta D,
\end{equation}
where $C_{\mathrm{SN}}$ is the covariance matrix of the supernova dataset. The residual vector $D$ is defined by
\begin{equation}
\Delta D_i = \mu_{\mathrm{obs}}(z_i) - \mu_{\mathrm{th}}(z_i),
\end{equation}
where $\mu_{\mathrm{obs}}(z_i)$ and $\mu_{\mathrm{th}}(z_i)$ denote the observed and theoretically predicted distance moduli of the $i$-th supernova at redshift $z_i$, respectively. This residual quantifies the deviation between observation and theory and is used to construct the likelihood, consistently accounting for observational uncertainties in parameter estimation.}

\textbf{DESI Data Release~2 (DR2) Dataset:}~We employ the BAO measurements from the DESI DR2~\cite{karim2025desi,Lodha2025DESI,bhagat2026logarithmic}. This dataset provides constraints in terms of either two correlated distance ratios, $D_M/r_d$ and $D_H/r_d$, 
or a single isotropic distance ratio, $D_V/r_d$, where $r_d$ denotes the comoving sound horizon at the drag epoch. The relevant distance measures are defined as
\begin{equation}
D_H(z) = \frac{c}{H(z)},
\end{equation}
\begin{equation}
D_M(z) = \frac{c}{H_0}\int_{0}^{z}\frac{dz'}{E(z')},
\end{equation}
and
\begin{equation}
D_V(z) = \left[z\,D_H(z)\,D_M^2(z)\right]^{1/3}.
\end{equation}
These measurements are obtained across multiple redshift bins using a variety of tracers, including bright galaxies (BGS), luminous red galaxies (LRG), emission-line galaxies (ELG), quasars (QSO), and the Lyman-$\alpha$ (Ly$\alpha$) forest observed in high-redshift quasars. Such a multi-tracer approach allows DESI to probe the BAO signal over a wide redshift range with high precision.

For the DESI BAO dataset, the statistical analysis is performed using the chi-square estimator
\begin{equation}
\chi^2_{\rm BAO} =
\sum_{i,j}
\left(D^{\rm obs}_i - D^{\rm th}_i\right)
(C^{-1})_{ij}
\left(D^{\rm obs}_j - D^{\rm th}_j\right),
\end{equation}
where $D^{\rm obs}_i$ and $D^{\rm th}_i$ denote the observed and model-predicted BAO distance indicators, respectively, and $(C^{-1})_{ij}$ is the inverse covariance matrix accounting for correlations between different measurements. The best-fit model parameters are determined by minimizing $\chi^2_{\rm BAO}$ with respect to the parameter space.

\textbf{Joint Likelihood and MCMC Inference:}~When several observational probes are combined, the overall statistical estimator is defined as the sum of the individual chi-square contributions. For the joint analysis of the {CC, Pantheon$^{+}$ and DESI DR2 datasets}, the total chi-square is given by
\begin{equation}\label{10:021}
\chi^2_{\rm CC+SN+BAO}(\boldsymbol{\theta}) =
\chi^2_{\rm CC}(\boldsymbol{\theta}) +
\chi^2_{\rm SN}(\boldsymbol{\theta})+
\chi^2_{\rm BAO}(\boldsymbol{\theta}).
\end{equation}
The total likelihood function, incorporating chi-squared terms, is sampled using the affine-invariant ensemble sampler from the `emcee' Python package. Prior distributions are assigned to each of the parameters, and the chains are evolved to explore the multidimensional posterior landscape. The final results yield marginalized posteriors, best-fit estimates, and credible intervals for all parameters.

\begin{figure}
\begin{center}
\includegraphics[height=7.0in]{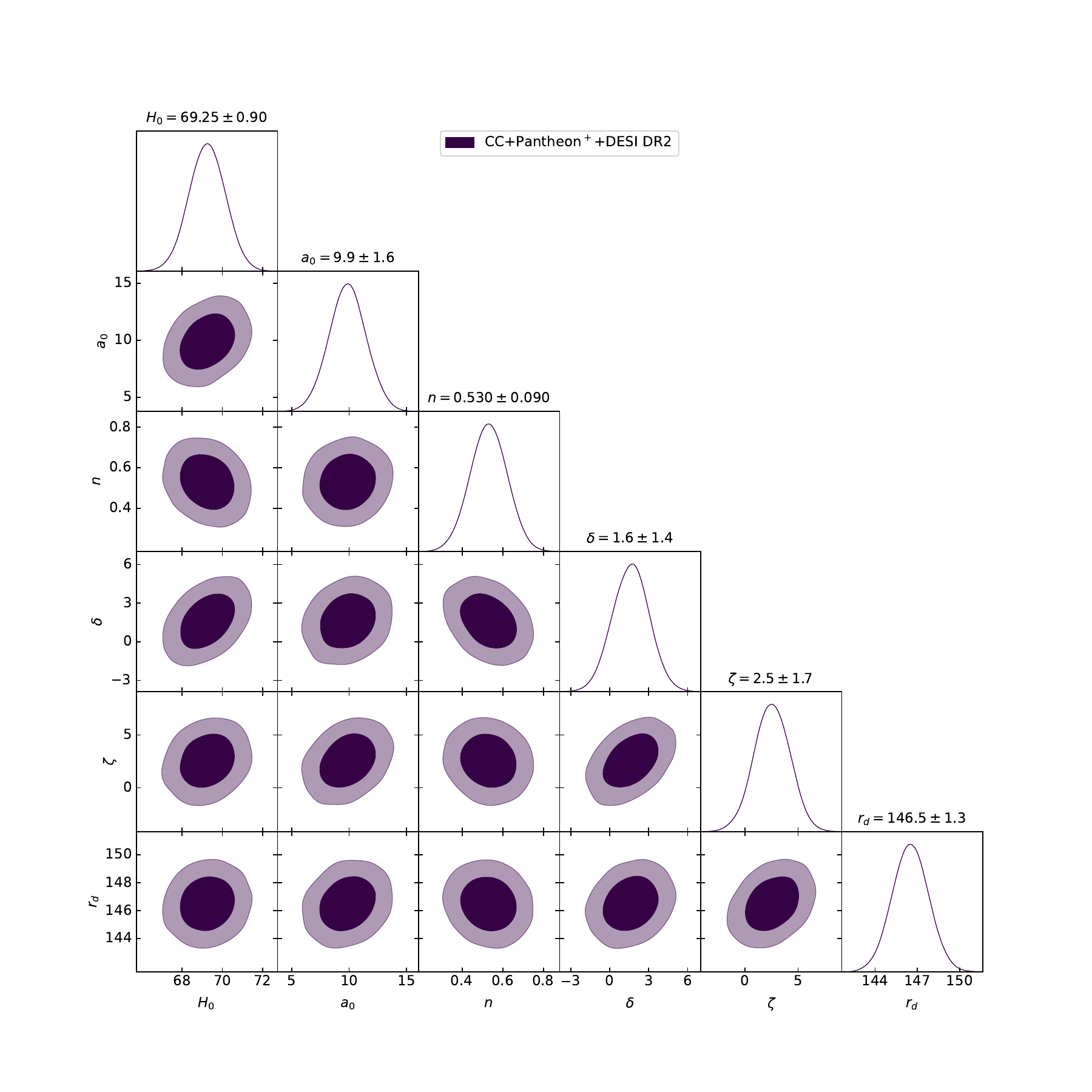}
\caption{{Two-dimensional marginalized posterior distributions for the cosmological parameters of the reconstructed model, derived from the combination of CC+Pantheon$^{+}$+DESI DR2 datasets.}}
\label{T}
\end{center}
\end{figure}

{\color{black}\subsubsection{Contour Analysis from Combined Observations}
Figure~\ref{T} shows the two-dimensional marginalized posterior distributions of the cosmological parameter space 
$\Theta=(H_0,a_0,n,\delta,\zeta,r_d)$ for the reconstructed model, obtained from the joint analysis of the CC, Pantheon$^{+}$, and DESI DR2 datasets. The contours correspond to the 68\% and 95\% confidence levels, representing the $1\sigma$ and $2\sigma$ credible regions, respectively, after marginalizing over the remaining parameters.

The diagonal panels display the one-dimensional marginalized posterior distributions for each parameter, from which the mean values and associated $1\sigma$ uncertainties are inferred. The Hubble constant $H_0$ is tightly constrained, reflecting the strong sensitivity of the combined datasets to the present expansion rate of the Universe. The parameters $a_0$ and $n$ are well bounded. In particular, the parameter $n$ governs the rate of cosmic expansion and therefore plays a crucial role in determining the late-time acceleration behavior, while $a_0$ fixes the normalization of the scale factor. Hence, it implies that the evolution of the scale factor has stable and monotonic behavior. This behavior ensures the absence of any sudden singularity in the evolutionary history. The Tsallis parameter $\delta$ quantifies deviations from standard Boltzmann--Gibbs entropy and controls the degree of non-extensivity in the dark energy sector, whereas $\zeta$ acts as a normalization constant for the holographic dark energy density. The marginalized posteriors indicate that both parameters are constrained within finite ranges, demonstrating that current observational data are capable of placing meaningful bounds on the non-extensive holographic dark energy dynamics. {The Planck $\Lambda$CDM CMB analysis predicts a sound horizon value $r_d \approx 147~\mathrm{Mpc}$. The posterior distribution of $r_d\approx 146.5\pm 1.3~\mathrm{Mpc}$ obtained in the present analysis (see Figure~\ref{T}) is consistent with its expected value within the corresponding uncertainties. Hence, Figure~\ref{T} also indicates compatibility of the reconstructed THDE-$f(Q,C)$ model with the physics of the early universe. BAO observations primarily constrained $H_0 r_d \approx  9.9 \times 10^3~\mathrm{kms^{-1}}$ \cite{karim2025desi,rd}. From our study, we have obtained $H_0$ and $r_d$ leading to $H_0 r_d \approx  1.01 \times 10^4~\mathrm{kms^{-1}}$ which falls within the constrained value of $H_0 r_d$.}

The off-diagonal panels illustrate the correlations between different parameter pairs. A mild degeneracy is observed between $H_0$ and $a_0$, reflecting their coupled influence on the background expansion rate. A mild correlation is observed between $a_0$ and $n$, consistent with their joint role in shaping the background expansion through the power-law scale factor. Correlations involving $\delta$ and $\zeta$ reflect their combined influence on the evolution of the Tsallis holographic dark energy density. Importantly, no strong degeneracies are present, and the confidence regions remain compact, indicating good convergence of the MCMC chains and statistical consistency of the reconstructed model.

Overall, the contour plot demonstrates that the reconstructed cosmological scenario in $f(Q,C)$ gravity, incorporating a power-law expansion and Tsallis holographic dark energy, provides a statistically viable and observationally consistent description of the late-time cosmic expansion when confronted with current high-precision datasets.}

{Following Fig.~\ref{T}, the constrained parameter values of the reconstructed model are presented in Table \ref{OD} using CC+Pantheon$^{+}$+DESI DR2 combined datasets \cite{CC4,karim2025desi,scolnic2022pantheonplus}. The first row of Table \ref{OD} represents the values of the cosmological parameters with $1\sigma$ uncertainties. In addition to these, the minimum chi-square value $\chi^{2}_{min}$ obtained for the best fit parameters is incorporated. The second row lists the flat (uniform) priors imposed on all free parameters used in the MCMC sampling, with the Hubble constant varying in the range $H_0 \in (60,90)\,$, the parameter $a_0 \in (0,20)$, the power-law exponent $n \in (0,5)$, the deformation parameter $\delta \in (0,3)$, the positive
constant $\zeta \in (0,5)$ and the comoving sound horizon $r_d \in (100,200)$. These indicate a broad exploration of the parameter space considered here. The resulting posterior distributions are then used to infer the best-fit values of the parameters together with their corresponding confidence intervals.}

\begin{figure}
\begin{center}
\includegraphics[height=5.0in,width=6.90in]{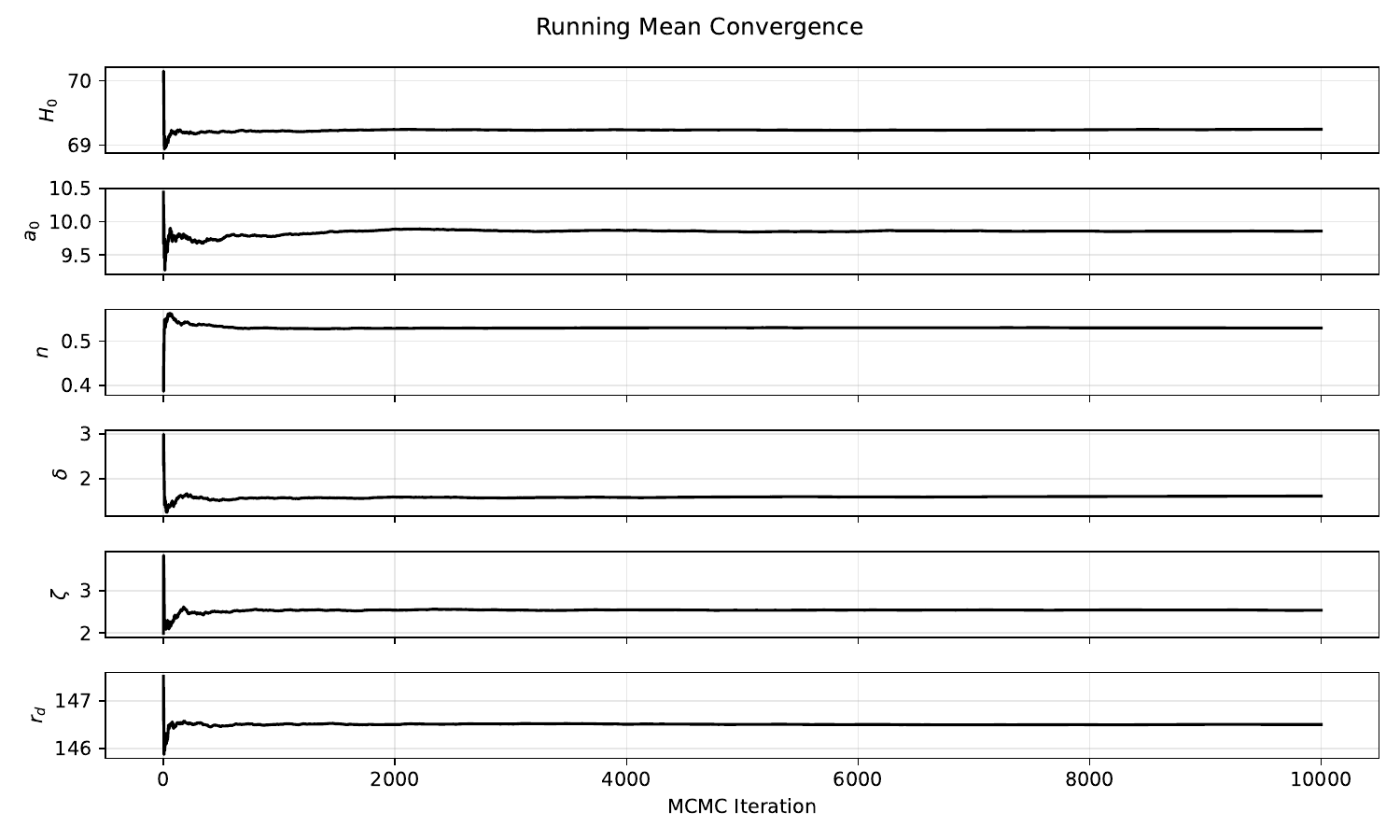}
\caption{{Running mean convergence of the MCMC chains for the cosmological parameters.}}
\label{T1}
\end{center}
\end{figure}

{In Fig.~\ref{T1}, we have illustrated the running mean convergence of the MCMC chains for the sampled cosmological parameters. It can be observed from Fig.~\ref{T1} that in the initial phase it is not stabilized, and gradually the running mean stabilizes and fluctuates around nearly constant values. From this behavior, we can interpret that the chains have reached a stationary distribution, confirming the convergence of the MCMC sampling \cite{Sampling}.}

\begin{figure}
\begin{center}
\includegraphics[height=3.5in]{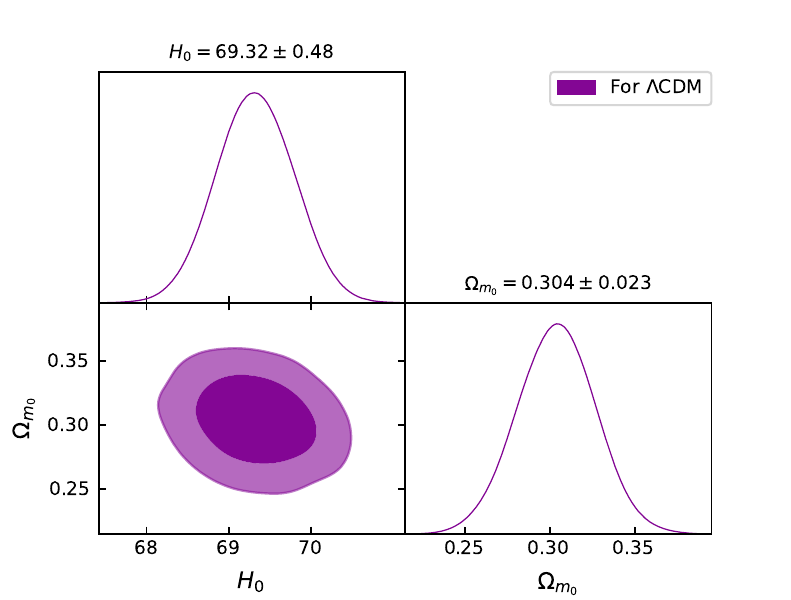}
\caption{{Two-dimensional marginalized posterior distributions for the cosmological parameters of the $\Lambda$CDM model, derived from the combination of CC+Pantheon$^{+}$+DESI DR2 datasets.}}
\label{LCDM}
\end{center}
\end{figure}

Fig.~\ref{LCDM} displays the two-dimensional marginalized posterior distributions of the cosmological parameters within the $\Lambda$CDM framework, obtained from the joint analysis of the CC, Pantheon$^{+}$, and DESI DR2 datasets. The confidence contours shown correspond to the 68\% and 95\% confidence levels. As shown in Fig.~\ref{LCDM}, Table~\ref{OD1} summarizes the constraints on the $\Lambda$CDM model parameters obtained from the joint analysis of the CC, Pantheon$^{+}$, and DESI DR2 datasets~\cite{CC4,karim2025desi,scolnic2022pantheonplus}. The first row of Table~\ref{OD1} reports the best-fit values of the cosmological parameters together with their corresponding $1\sigma$ confidence intervals. In addition, the minimum chi-square value, $\chi^2_{\min}$, associated with the optimal fit is also included. The second row lists the flat (uniform) priors, with the Hubble constant varying in the range $H_0 \in (60,90)\,$ and the current matter density $\Omega_{m_0} \in (0,0.5)$. {The difference in the minimum chi-square values between the reconstructed model and the standard $\Lambda$CDM model is $\Delta\chi^{2}_{min}=2.041$. According to the minimum $\chi^2$ criterion, this indicates that $\Lambda$CDM provides a slightly better fit. However, the reconstructed THDE-$f(Q,C)$ framework yields a fit to the observational data comparable to that of $\Lambda$CDM. This suggests that the model remains observationally viable within the considered parameter space.} Hence, in general, Table~\ref{OD} shows that the THDE-$f(Q,C)$ model provides a good fit to the observations, and it also indicates the non-extensive holographic dark energy as a viable extension of $\Lambda$CDM. 

\begin{table}[ht!]
\caption{{Constrained parameter values of the reconstructed model for CC+Pantheon$^{+}$+DESI DR2 combined datasets \cite{CC4,karim2025desi,scolnic2022pantheonplus}.}}
\centering
 \begin{tabular}{||c | c | c | c | c | c | c | c ||} 
 \hline
~~~~~~Data set~~~~~~ & ~~~~~~$H_0$~~~~~~ & ~~~~~~$a_0$~~~~~~&~~~~~~$n$~~~~~~&~~~~~~$\delta$~~~~~~&~~~~~~$\zeta$~~~~~~&~~~~~~$r_d$~~~~~~&~~~~~~$\chi^2_{min}$~~~~~~\\ [1ex] 
 \hline\hline
 CC+Pantheon$^+$+DESI DR2 & $69.25\pm 0.90$ & \(9.9\pm 1.6\) & \(0.530\pm 0.090\) & \(1.6\pm 1.4\) & \(2.5\pm 1.7\) & $146.5\pm 1.3$ & 1661.357\\[1ex]
\hline
 Prior Ranges & (60,90)& (0,20) & (0,5) & (0,3) & (0,5)& (100,200) & - \\[1ex]
 \hline\hline
 \end{tabular}
 \label{OD}
\end{table}

\begin{table}[ht!]
\caption{{Constrained parameter values of the $\Lambda$CDM model for CC+Pantheon$^{+}$+DESI DR2 combined datasets \cite{CC4,karim2025desi,scolnic2022pantheonplus}.}}
\centering
 \begin{tabular}{||c | c | c | c  ||} 
 \hline
~~~~~~Data set~~~~~~ & ~~~~~~$H_0$~~~~~~ & ~~~~~~$\Omega_{m_0}$~~~~~~&~~~~~~$\chi^2_{min}$~~~~~~\\ [1ex] 
 \hline\hline
 CC+Pantheon$^+$+DESI DR2 & $69.32\pm 0.48$ & \(0.304\pm 0.023\) & 1659.316\\[1ex]
\hline
 Prior Ranges & (60,90)& (0,0.5) & - \\[1ex]
 \hline\hline
 \end{tabular}
 \label{OD1}
\end{table}

\begin{figure}
\begin{center}
\includegraphics[height=3.5in]{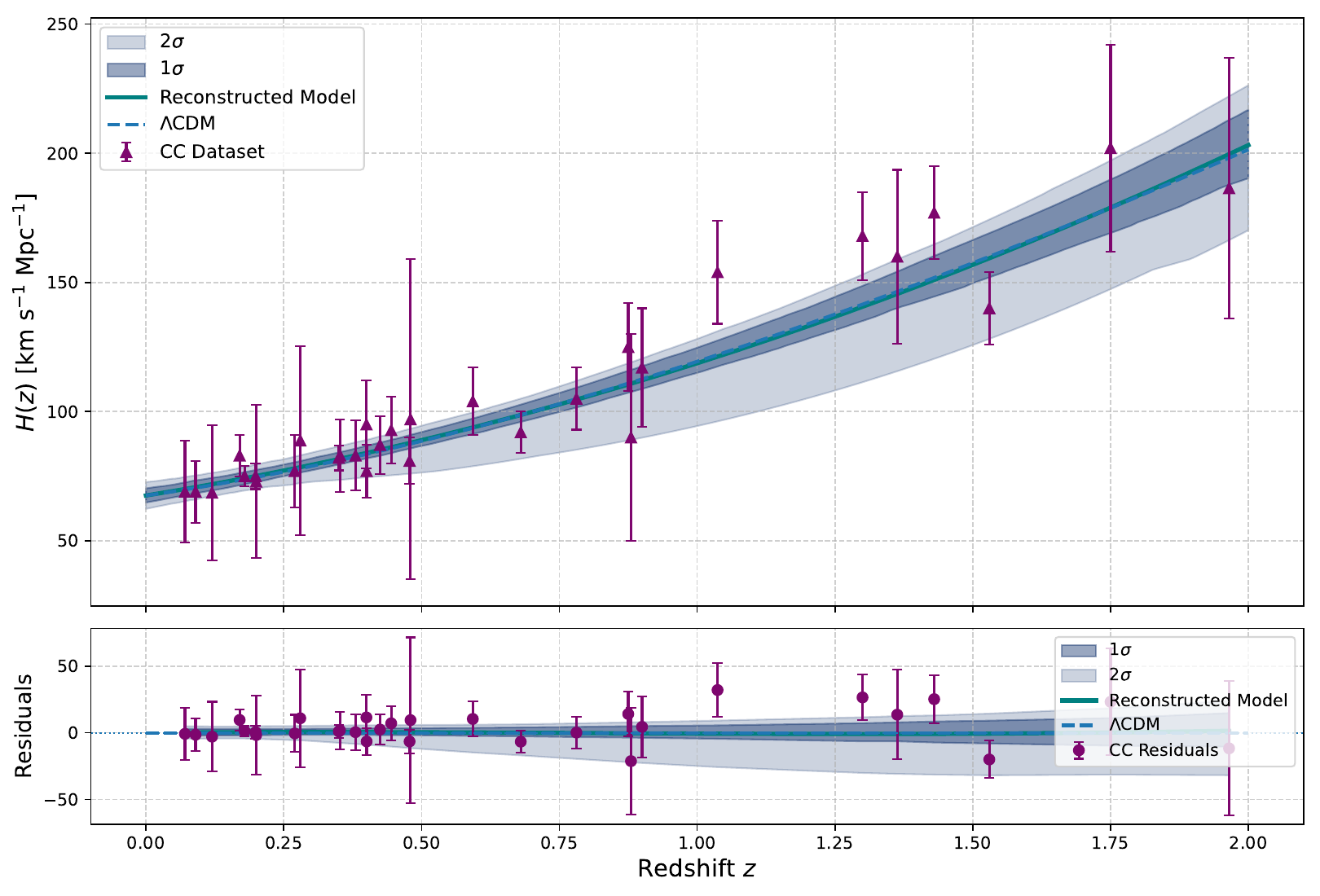}
\caption{{Reconstructed Hubble parameter $H(z)$ for the THDE model in the $f(Q,C)$ framework (upper panel), constrained by the combined datasets. The best-fit theoretical curve (solid line) is shown together with the CC measurements and their corresponding uncertainties. The $\Lambda$CDM prediction based on its best-fit parameters (dashed line) is included for comparison. The lower panel displays the residuals as a function of redshift, with error bars indicating standard uncertainties.}}
\label{H}
\end{center}
\end{figure}

{\subsubsection{Comparison of reconstructed $H(z)$ against CC data set}
In Fig.~\ref{H}, the reconstructed Hubble parameter $H(z)$ constrained using the combined datasets is plotted with respect to redshift $z$ for the reconstructed THDE in the $f(Q,C)$ framework. For comparison, the standard $\Lambda$CDM is plotted as the dashed curve, whereas the solid curve represents $H(z)$ from the reconstruction scheme. Furthermore, to assess the behavior using observational data, we have the observational data points corresponding to CC measurements, along with the error bars represented by the arrowed vertical lines. The shaded regions denote the $1\sigma$ and $2\sigma$ confidence intervals around the reconstructed model. We observe from Fig.~\ref{H} that the reconstructed THDE-$f(Q,C)$ model stays within the observational bounds for the entire range of redshift under consideration. Throughout the range of redshift, as observed in the upper panel, the THDE-$f(Q,C)$ model is close to $\Lambda$CDM, and it is also noted that the widths of the confidence intervals have significantly reduced from higher to lower redshift. It is noted that at lower redshifts, $1\sigma$ and $2\sigma$ confidence intervals are almost coincident. Hence, it may be interpreted that at lower redshifts, the degrees of uncertainty are significantly reduced, indicating better precision. In the lower panel of Fig.~\ref{H}, we present the residuals of the Hubble parameter $H(z)$ computed using the 32-point CC dataset, in order to assess the agreement between the reconstructed cosmological model and the observational measurements. For each redshift bin $z_i$, the residual is defined as $\Delta H(z_i) = H_{\rm obs}(z_i) - H_{\rm th}(z_i)$, where $H_{\rm obs}(z_i)$ denotes the observed Hubble parameter from the CC compilation and $H_{\rm th}(z_i)$ represents the corresponding theoretical prediction evaluated using the combined datasets of best-fit model parameters. The residuals fluctuate around $0$ and are contained mainly within the confidence band. This indicates statistical consistency of the reconstructed model with the CC observations. Combining the outcomes of the two panels of Fig.~\ref{H}, we can say that the THDE-$f(Q,C)$ framework successfully reproduces the observed evolutionary behavior of $H(z)$. The compatibility with the current observations is understandable from the close agreement of the model with $\Lambda$CDM. Consequently, the reconstructed model provides a theoretically viable and observationally consistent alternative explanation of the cosmic expansion.}

\begin{figure}
\begin{center}
\includegraphics[height=3.5in]{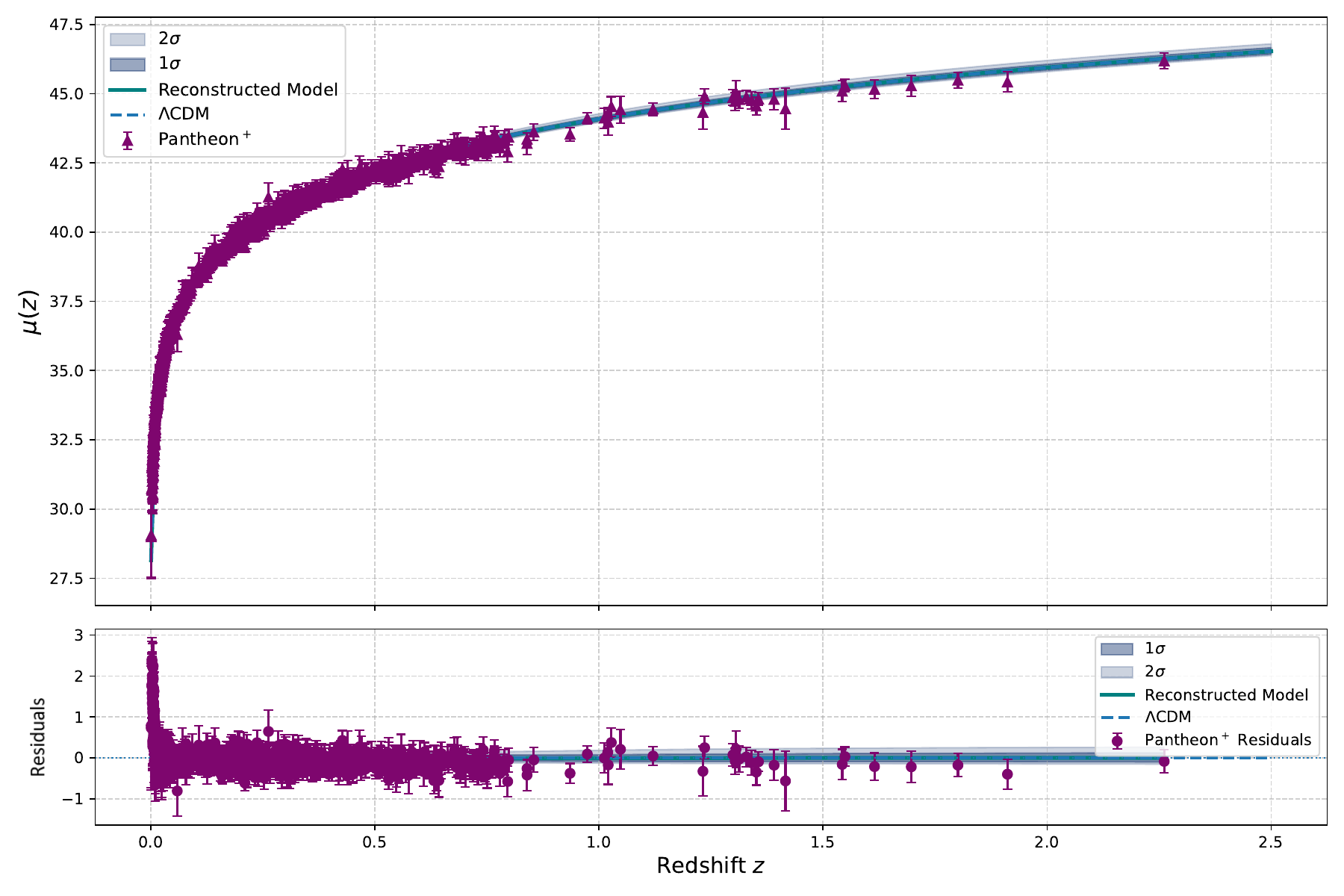}
\caption{{Plot of $\mu(z)$ for the THDE model reconstructed in the framework of $f(Q,C)$ gravity (upper panel), constrained by the combined datasets. The theoretical prediction (solid line) is compared against the observational data from the Pantheon$^{+}$ dataset. The $\Lambda$CDM prediction based on its best-fit parameters (dashed line) is included for comparison. The lower panel shows the residuals as a function of redshift, highlighting deviations from the observed values. The error bars represent the corresponding standard uncertainties.}}
\label{P}
\end{center}
\end{figure}

{\subsubsection{Comparison of $\mu(z)$ for the Reconstructed THDE-$f(Q,C)$ Model and Pantheon$^{+}$ Dataset}
The Fig.~\ref{P} shows the distance modulus for the reconstructed THDE in $f(Q,C)$ gravity and the $\Lambda$CDM model constrained using the combined datasets, and a comparison is presented with the Pantheon$^{+}$ data point set. In the upper panel, the solid curve represents the best fit reconstructed $\mu(z)$ obtained from the  THDE-$f(Q,C)$ model. On the other hand, the dashed curve pertains to the $\Lambda$CDM model. The shaded region indicates $1\sigma$ and $2\sigma$ confidence intervals around the reconstructed model. The Pantheon$^{+}$ data points with error bars are also presented. The reconstructed model exhibits close agreement with the Pantheon$^{+}$ dataset over the entire redshift range. The closeness implies consistency with the observational data. The lower panel of Fig.~\ref{P} illustrates the residuals of the distance modulus $\mu(z)$ for the 1701-point Pantheon$^{+}$ supernova sample, providing a quantitative assessment of how accurately the model reproduces the observed luminosity--distance relation. The residual at each redshift $z_i$ is defined as $\Delta \mu(z_i) = \mu_{\rm obs}(z_i) - \mu_{\rm th}(z_i)$, where $\mu_{\rm obs}(z_i)$ corresponds to the measured distance modulus from the Pantheon$^{+}$ catalogue, while $\mu_{\rm th}(z_i)$ denotes the model prediction evaluated using best-fit parameter set obtained from the combined analyses. The residuals are distributed about $0$ and are mostly within the shaded confidence region. This implies statistical consistency of the model with the Pantheon$^{+}$ dataset. Hence, this reconfirms that the reconstructed model is observationally viable.}

\subsection{Deceleration and EoS parameters}

\begin{figure}
\begin{center}
\includegraphics[height=3.5in]{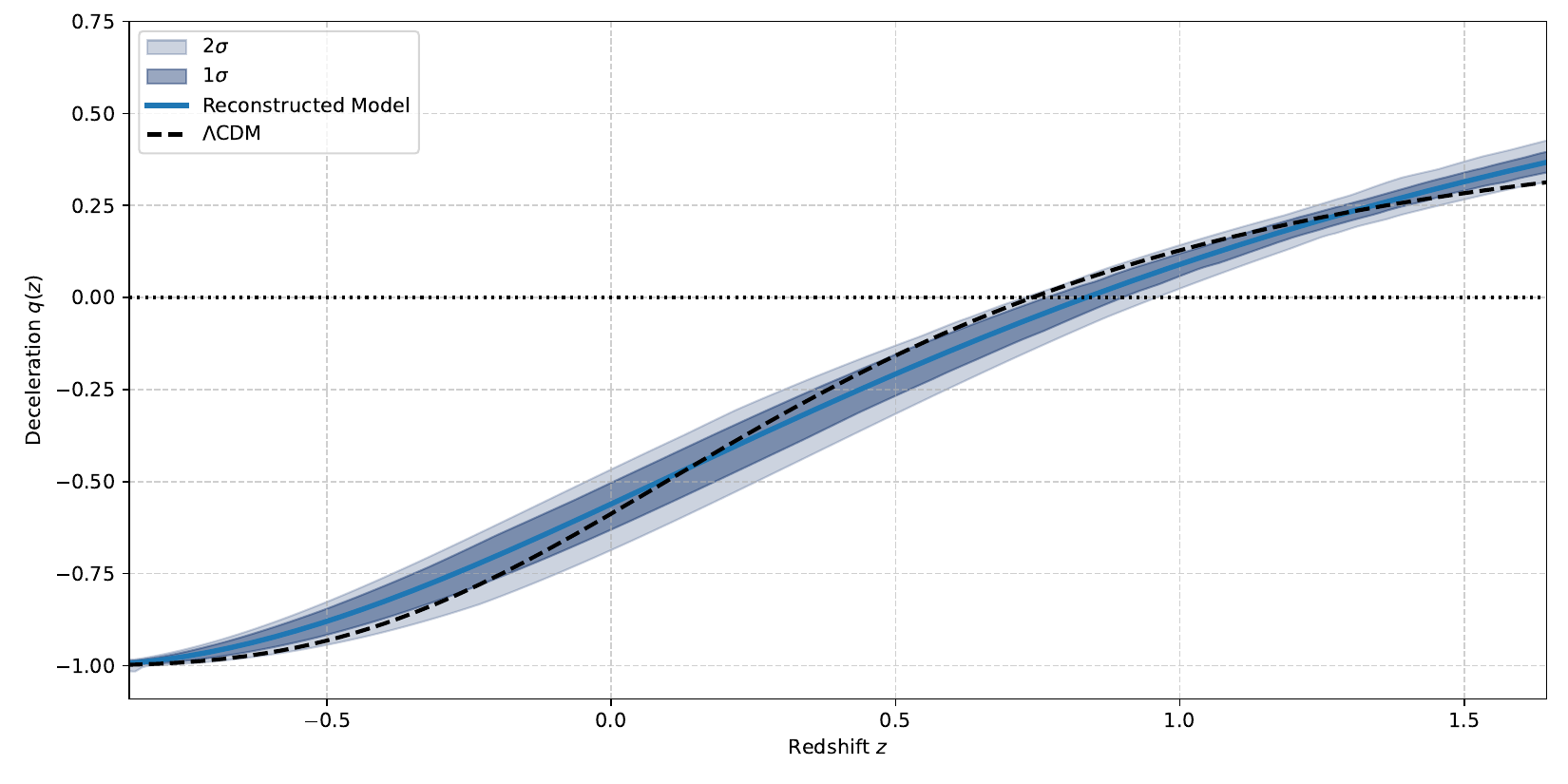}
\caption{{Redshift evolution of the deceleration parameter $q(z)$ for THDE model reconstructed in the framework of $f(Q,C)$ gravity, compared with the $\Lambda$CDM prediction.}}
\label{q}
\end{center}
\end{figure}

In addition to analyzing the Hubble parameter, we also examined other cosmological parameters to validate and further characterize our model. To demonstrate the current accelerated expansion of the Universe, we analyzed the deceleration parameter (solid curve) $q(z)=-1+\frac{(1+z)}{H(z)}\frac{dH(z)}{dz}$ with $1\sigma$ and $2\sigma$ confidence intervals and plotted it in Fig.~\ref{q}. From Fig.~\ref{q}, we can see that during the early phase of the Universe, specifically for redshifts \( z > 0.75 \), the deceleration parameter \( q(z) \) is greater than zero, indicating a decelerated expansion phase. At approximately \( z_t \approx 0.8 \), there is a notable transition where the deceleration parameter \( q(z) \) shifts from a positive value to a negative value for the combined data sets CC+Pantheon$^{+}$+DESI DR2. This change signifies that the Universe is moving from a decelerated expansion phase (where \( q(z) > 0 \)) to an accelerated expansion phase (where \( q(z) < 0 \)). Eventually, the deceleration parameter converges to \(-1\) and approaches this value asymptotically. The $\Lambda$CDM prediction (dashed curve) is included as a reference, with the reconstructed model closely matching its behavior while exhibiting mild deviations at intermediate redshifts indicative of dynamical dark energy effects. Therefore, we conclude that it is indeed possible for the Universe to transition from a decelerated expansion phase in its early stages to an accelerated expansion phase at later times for THDE in $f(Q,C)$ gravity.

\begin{figure}
\begin{center}
\includegraphics[height=3.5in]{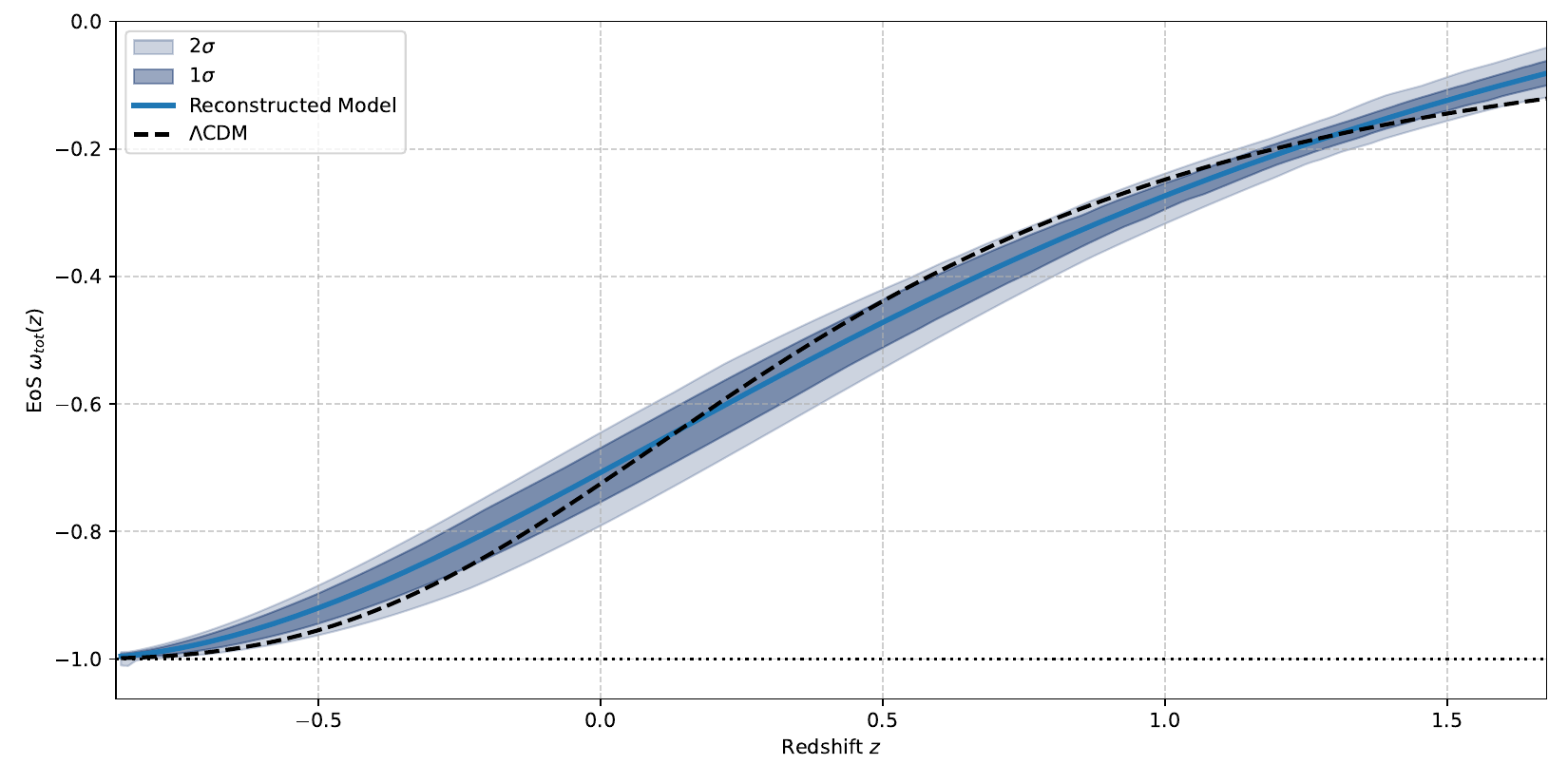}
\caption{{Redshift evolution of the EoS parameter ${\omega_{tot}(z)}$ for THDE model reconstructed in the framework of $f(Q,C)$ gravity, compared with the $\Lambda$CDM prediction.}}
\label{w}
\end{center}
\end{figure}

We have also examined the behavior of the EoS parameter (solid curve) ${\omega_{tot}(z)}=-1+\frac{2(1+z)}{3H(z)}\frac{dH(z)}{dz}$ with $1\sigma$ and $2\sigma$ confidence intervals in Fig.~\ref{w}. In the very early Universe, the EoS parameter, which relates pressure to energy density, can take on a positive value. This indicates that the Universe's pressure is proportional to its energy density. Depending on the specific value of the EoS parameter, this relationship could lead to scenarios such as rapid expansion or even contraction of the Universe. However, with the evolution of the Universe, the EoS parameter has gradually entered into negative regime satisfying ${\omega_{tot}(z)}<-\frac{1}{3}$ as required by the accelerated expansion of the Universe. For THDE in $f(Q,C)$ gravity, we have plotted the reconstructed EoS parameter based on the reconstructed pressure and density, as already elaborated in the previous section. Choosing the values of the parameters constrained through MCMC, we have plotted its evolution in Fig.~\ref{w} for late time, and it is observed that it shows quintessence behavior, and for $z=0$, it tends to \(-1\) and gradually becomes asymptotic to it at a later stage. Notably, the crossing of the phantom boundary is not available here. For reference, the $\Lambda$CDM prediction (dashed curve) is included. The reconstructed model reproduces its overall behavior, with mild deviations at intermediate redshifts reflecting dynamical dark energy effects. 

Hence, the negative value of the deceleration parameter, nearing \(-1\), along with the behavior of the EoS parameter, indicates that the Universe is currently undergoing acceleration, which aligns with the predictions of the $\Lambda$CDM model. The current values of $q(z)$ and ${\omega_{tot}(z)}$ constrained using the combined datasets CC+Pantheon$^{+}$+DESI DR2 for reconstructed THDE in $f(Q,C)$ gravity are presented in Table \ref{CV}.

\begin{table}[ht!]
\caption{{Current values of Cosmological parameters for THDE model reconstructed in the framework of $f(Q,C)$ gravity for the combined datasets CC+Pantheon$^{+}$+DESI DR2 \cite{CC4,karim2025desi,scolnic2022pantheonplus}.}}
\centering
 \begin{tabular}{||c | c ||} 
 \hline
~~~~~~Cosmological parameters~~~~~~ & ~~~~~~CC+Pantheon$^{+}$+DESI DR2~~~~~~\\ [1ex] 
 \hline\hline
 $q_0$ & $-0.561^{+0.058}_{-0.070}$\\[1ex]
 \hline
 ${\omega_{tot,0}}$ & $-0.707^{+0.038}_{-0.046}$\\[1ex]
 \hline
 $j_0$ & $1.338^{+0.032}_{-0.038}$\\[1ex]
 \hline
 $s_0$ & $-0.428^{+0.075}_{-0.064}$\\[1ex]
 \hline\hline
 \end{tabular}
 \label{CV}
\end{table}

\subsection{Diagnostics}
To better understand and distinguish between different cosmological models, especially in the context of dark energy, diagnostics have been proposed as a powerful geometrical tool.
{\color{black} \subsubsection{Jerk parameter}
In cosmology, the jerk parameter \cite{visser2004jerk,mukherjee2016parametric,mukherjee2021non,dunajski2008cosmic} is a dimensionless quantity that characterizes the third derivative of the cosmic scale factor $a(t)$ with respect to cosmic time $t$ and provides a refined description of the Universe’s expansion beyond the Hubble and deceleration parameters. It is defined as
\begin{equation}
j \equiv \frac{1}{a H^{3}} \frac{d^{3}a}{dt^{3}},
\end{equation}
where the time derivatives can be expressed in terms of the redshift $z$ using the relation $1+z=1/a$. In the standard $\Lambda$CDM framework, the jerk parameter remains constant with $j=1$ at all redshifts, making it a powerful null diagnostic for testing departures from the concordance model. {As a diagnostic test, we analyzed the redshift evolution of the jerk parameter $j(z)$ (solid curve) with $1\sigma$ and $2\sigma$ confidence bands in Fig.~\ref{jerk} and compared it with the $\Lambda$CDM (dashed curve) benchmark value $j=1$. Since the jerk parameter serves as a purely geometrical indicator of the expansion history, deviations from $j=1$ provide a sensitive probe of departures from the standard $\Lambda$CDM scenario. It may be noted that at intermediate redshift ($z>0$), mild to moderate deviations from $j=1$. It may have appeared depending on the choice of parameters in the reconstruction scheme. However, the model reproduces the concordance behavior at the late time limit and remains consistent with the combined observational datasets (CC+Pantheon$^{+}$+DESI DR2). A detailed assessment of the evolutionary behavior of the universe at very high redshift, particularly around the CMB epoch ($z \approx 1100$), would require the inclusion of CMB likelihood data. Such an investigation is beyond the scope of the current study and may be explored in future studies.}

\begin{figure}
\begin{center}
\includegraphics[height=3.5in]{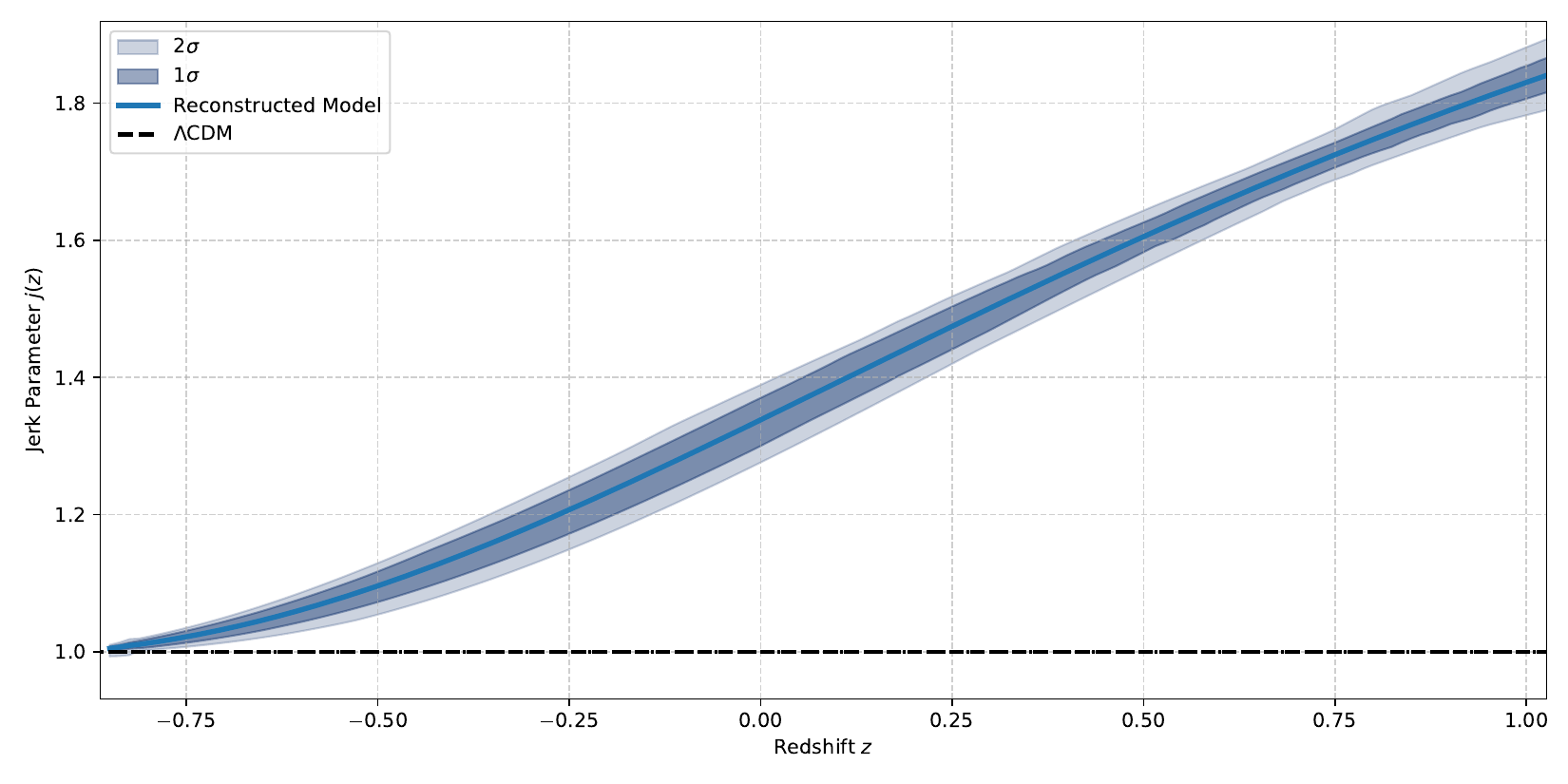}
\caption{{Redshift evolution of the jerk parameter $j(z)$ for THDE model reconstructed in the framework of $f(Q,C)$ gravity, compared with the $\Lambda$CDM prediction.}}
\label{jerk}
\end{center}
\end{figure}

\subsubsection{Snap parameter}
The snap parameter \cite{visser2004jerk,dunajski2008cosmic}, associated with the fourth derivative of the scale factor, extends the cosmographic description to even higher order and captures finer details of the cosmic expansion history. It is defined as
\begin{equation}
s \equiv \frac{1}{a H^{4}} \frac{d^{4}a}{dt^{4}},
\end{equation}
and can likewise be reformulated in terms of redshift derivatives of the Hubble parameter or the jerk parameter. {We analyzed the redshift evolution of the snap parameter $s(z)$, represented by the solid curve, along with $1\sigma$ and $2\sigma$ confidence bands shown in Fig.~\ref{snap}. This analysis was compared to the $\Lambda$CDM model, depicted by the dashed curve. It is important to distinguish between the present epoch ($z=0$) and the asymptotic future ($z\to -1$). In the standard flat $\Lambda$CDM cosmology, the snap parameter at the present epoch is finite and depends on the matter density parameter $\Omega_{m_0}$. For $\Omega_{m_0}=0.3$, $S_0=-0.35$ \cite{Jesus}. In the present case, the snap parameter at $z=0$ comes out to be $S_0=-0.428^{+0.075}_{-0.064}$, which is in close agreement with \cite{Jesus}. Therefore, the snap parameter does not approach zero at the present epoch. In this study, the cosmographic quantities are discussed primarily within the observed redshift range. The behavior for $z<0$ corresponds solely to the theoretical future evolution of the model. It represents a theoretical behavior and is not intended to draw any observationally supported conclusion.} The current values of $j(z)$ and $s(z)$ constrained using the combined datasets CC+Pantheon$^{+}$+DESI DR2 for reconstructed THDE in $f(Q,C)$ gravity are presented in Table \ref{CV}.

\begin{figure}
\begin{center}
\includegraphics[height=3.5in]{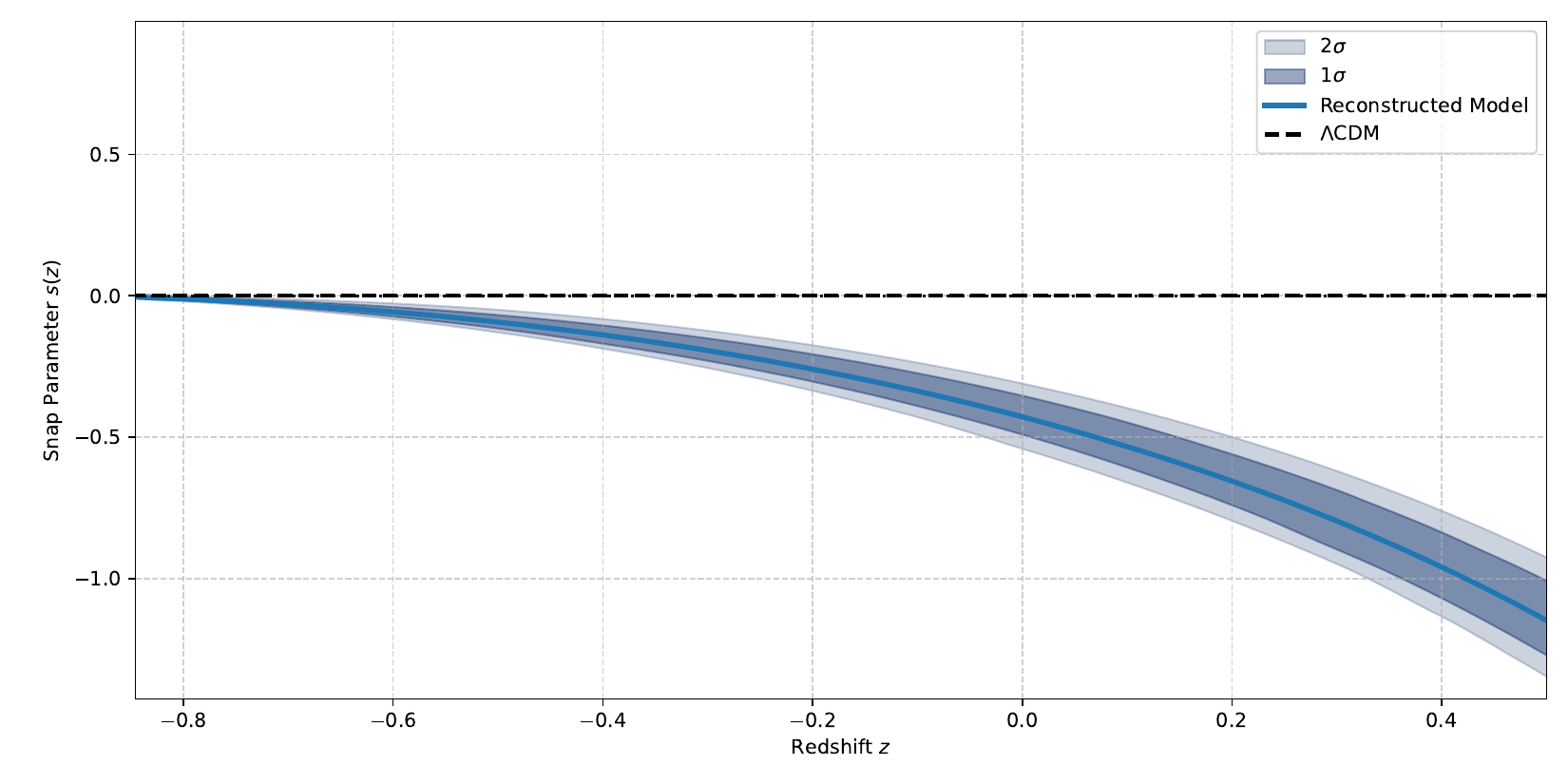}
\caption{{Redshift evolution of the snap parameter $s(z)$ for THDE model reconstructed in the framework of $f(Q,C)$ gravity, compared with the $\Lambda$CDM prediction.}}
\label{snap}
\end{center}
\end{figure}

\subsubsection{Statefinder pairs}
The Statefinder diagnostic \((r, s)\) \cite{S8,S9} is constructed from higher derivatives of the scale factor and provides a model-independent means to compare theoretical predictions with observational data. The cosmological diagnostic pair is directly related to the scale factor \(a\) and, consequently, to the metric that describes space-time. For this reason, it is considered “geometrical.” The parameters \(r\) and \(s\) are defined in terms of the Hubble parameter \(H\) and its time derivatives as
\begin{equation}
r=1+3\frac{\dot{H}}{H^{2}}+\frac{\ddot{H}}{H^{3}}
    \label{r1}
\end{equation}
and
\begin{equation}
s=-\frac{3H\dot{H}+\ddot{H}}{3H(2\dot{H}+3H^{2})}
    \label{r2}
\end{equation}
respectively. Different DE models \cite{S8,S9} represent different combinations of $r$ and $s$, as presented in Table \ref{tab:statefinder_models_bordered}.
\begin{table}[ht!]
\centering
\renewcommand{\arraystretch}{1.4}
\caption{Statefinder parameter values corresponding to different cosmological models.}
\label{tab:statefinder_models_bordered}

\begin{tabular}{|c|c|c|}
\hline
~~~~~~~~~~~~~~~~\textbf{$r$}~~~~~~~~~~~~~~~~ & ~~~~~~~~~~~~~~~~\textbf{$s$}~~~~~~~~~~~~~~~~ & ~~~~~~~~~\textbf{Corresponding Model} ~~~~~~~~~\\
\hline
$1$ & $0$ & $\Lambda$CDM model \\
\hline
$>1$ & $<0$ & Chaplygin Gas (CG) model \\
\hline
$1$ & $1$ & Standard Cold Dark Matter (SCDM) model \\
\hline
$<1$ & $>0$ & Quintessence region \\
\hline
$1$ & $\frac{2}{3}$ & Holographic Dark Energy (HDE) model \\
\hline
\end{tabular}

\end{table}

The evolutionary paths of the HDE model in the \( s-r \) plane \cite{tHooft1993dimensional,T12,T1,li2004model,14,15,16}, with the future event horizon serving as the infrared cut-off, start at the point \( s = \frac{2}{3}, r = 1 \) and eventually approach the \(\Lambda\)CDM fixed point at \( (s = 0, r = 1) \) over time \cite{15}. For both the Ricci dark energy (RDE) model and the quintessence dark energy model with a constant equation of state EoS parameter \cite{S8,S9}, the trajectories in the \( s-r \) plane are vertical \cite{17}. In the case of Chaplygin gas (CG), the trajectory remains in the regions where \( s < 0 \) and \( r > 1 \) \cite{18}. Conversely, the phantom model with a power-law potential and the quintessence models (inverse power-law) occupy the regions where \( s > 0 \) and \( r < 1 \)  \cite{S8,S9}, and both of these scenarios also approach the \(\Lambda\)CDM fixed point over time. In coupled quintessence models \cite{19}, the trajectory in the \( s-r \) plane exhibits a swirling pattern before reaching the attractor. During the early phase of the Universe, \(\Lambda\)CDM behavior is evident in both the Agegraphic dark energy model \cite{20} and the Polytropic gas model \cite{21}.
 The ghost dark energy (DE) model and the HDE model with a model parameter \( c = 1 \) exhibit similar behavior in the \((s, r)\) plane \cite{22}. This behavior is also consistent with other DE models, such as the generalized Chaplygin gas \cite{23,24,25}, Chaplygin gas \cite{26,27}, Yang-Mills \cite{28}, new agegraphic \cite{20,29}, and HDE \cite{14,15,16}. In the case of the tachyon DE model \cite{30} and the HDE model with the Granda–Oliveros infrared (IR) cut-off \cite{31}, the curve in the \((s, r)\) plane intersects the \(\Lambda\)CDM fixed point at a midpoint in the Universe's evolution. For the THDE model, the trajectories in the \((s, r)\) plane converge at the \(\Lambda\)CDM fixed point \((s = 0, r = 1)\) at late times, starting from the matter-dominated (SCDM) point \((s = 1, r = 1)\) and traversing an arc segment followed by a downward parabola \cite{tavayef2018tsallis,33}. The Chaplygin gas behavior in the RHDE model \cite{34,034} is illustrated by an evolutionary curve in the \((s, r)\) plane that starts and finishes at a swirl around the \(\Lambda\)CDM fixed point \((s = 0, r = 1)\). Recent research by one of the authors indicates that during the late-time evolution of the Universe, the Statefinder pair \((r,s)\) of the SMHDE model approaches the \(\Lambda\)CDM fixed point \((r = 1, s = 0)\) and consistently lies within the Chaplygin gas region \cite{35}. The evolution of the \((r,s)\) pair for the NGCG model has been analyzed in \cite{36}. For the Tsallis agegraphic dark energy model, the evolutionary curve in the \((s, r)\) plane begins at a cosmological constant, makes a circular turn, and then moves toward a different endpoint \cite{37}. Additionally, the authors of \cite{38,40} have provided a comprehensive analysis of the Statefinder pair for various DE models.
 
 The evolutionary trajectory of the Statefinder pair \((r,s)\) for reconstructed THDE in \(f(Q,C)\) gravity constrained using the combined datasets CC+Pantheon$^{+}$+DESI DR2 is illustrated in Fig.~\ref{rs}. From this figure, we observe that the trajectory of the reconstructed Statefinder pair resides in the quintessence region i.e. \((r<1,s>0)\), and passes through the \(\Lambda\)CDM fixed point, which is at \((r=1,s=0)\). The fact that the trajectory crosses the \(\Lambda\)CDM fixed point strongly indicates that the THDE reconstructed in \(f(Q,C)\) gravity revolves around the \(\Lambda\)CDM phase of the Universe. 
 
\begin{figure}
\begin{center}
\includegraphics[height=3.5in]{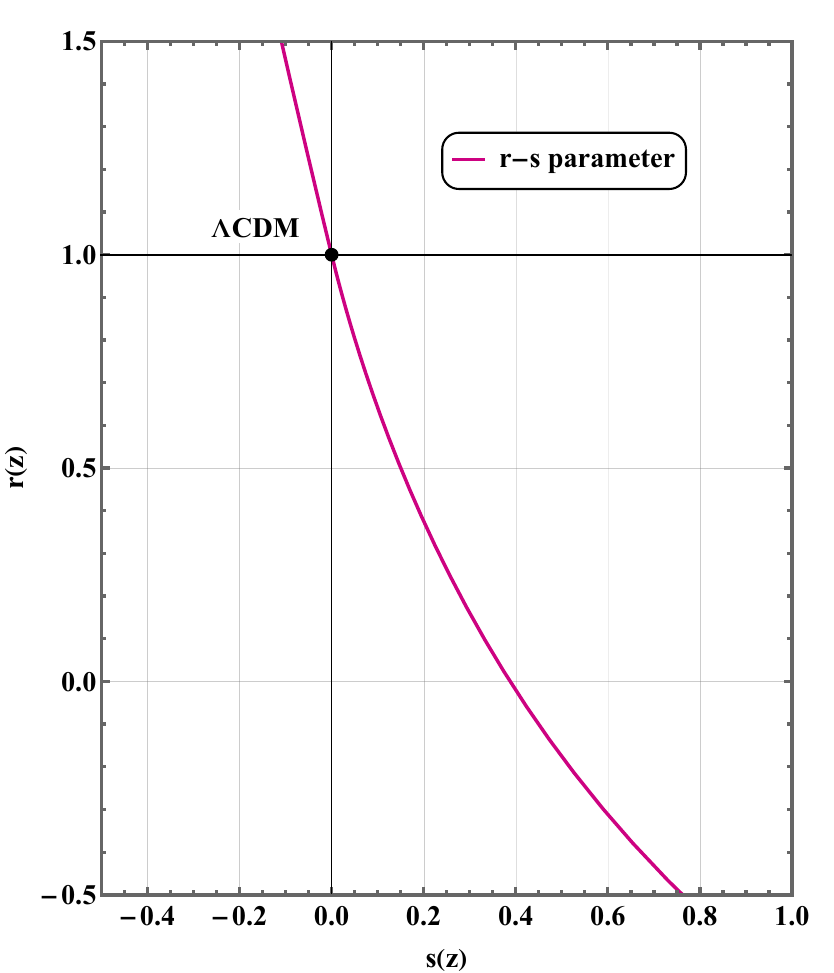}
\caption{{The statefinder pair $(r,s)$ trajectory for THDE model reconstructed in the framework of $f(Q,C)$ gravity.}}
\label{rs}
\end{center}
\end{figure}

The evolutionary trajectory of another Statefinder pair \((r, q)\) for reconstructed THDE in \(f(Q,C)\) gravity constrained using the combined datasets CC+Pantheon$^{+}$+DESI DR2 is illustrated in Fig.~\ref{rq}. In this representation, different cosmological epochs occupy distinct regions of the $(r,q)$ space. The radiation-dominated era corresponds to $(q,r)=(1,3)$, while the matter-dominated phase is characterized by $(q,r)=(0.5,1)$. The $\Lambda$CDM model is represented by the fixed point $(q,r)=(-1,1)$, which is marked explicitly in the figure and serves as a reference benchmark. The reconstructed trajectory begins at positive values of the deceleration parameter, indicating a decelerating expansion consistent with a matter-dominated Universe. As cosmic time evolves, the trajectory smoothly transitions toward negative values of $q$, signaling the onset of late-time acceleration. Simultaneously, $r\simeq 1$, demonstrating convergence toward the $\Lambda$CDM fixed point at low redshifts. Overall, the $(r,q)$ trajectory confirms that the reconstructed model successfully interpolates between the standard matter-dominated era and the late-time accelerated phase.
\begin{figure}
\begin{center}
\includegraphics[height=3.5in]{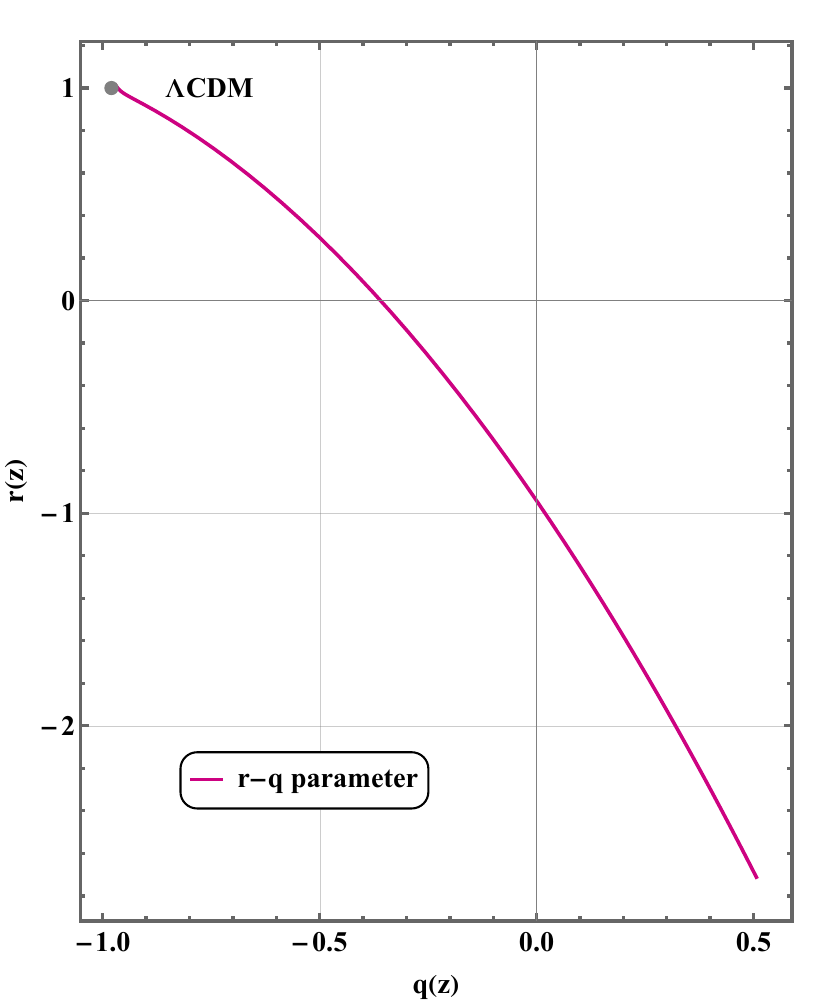}
\caption{{The statefinder pair $(r,q)$ trajectory for THDE model reconstructed in the framework of $f(Q,C)$ gravity.}}
\label{rq}
\end{center}
\end{figure}

\subsection{Age of the Universe}

We know that the rate of expansion of the Universe is determined by the Hubble parameter. The Hubble constant is a key observable in cosmology. The age of the Universe $t_0$ is estimated  through
\begin{equation}
    t_0 = \int_0^{\infty} \frac{dz}{(1+z)H(z)}.
\end{equation}
This relation links the expansion of the Universe to its overall age \cite{Bhagat2023, aghanim2020planck}. Hence, it provides a fundamental consistency check for cosmological models.

\begin{figure}
\begin{center}
\includegraphics[height=3.5in]{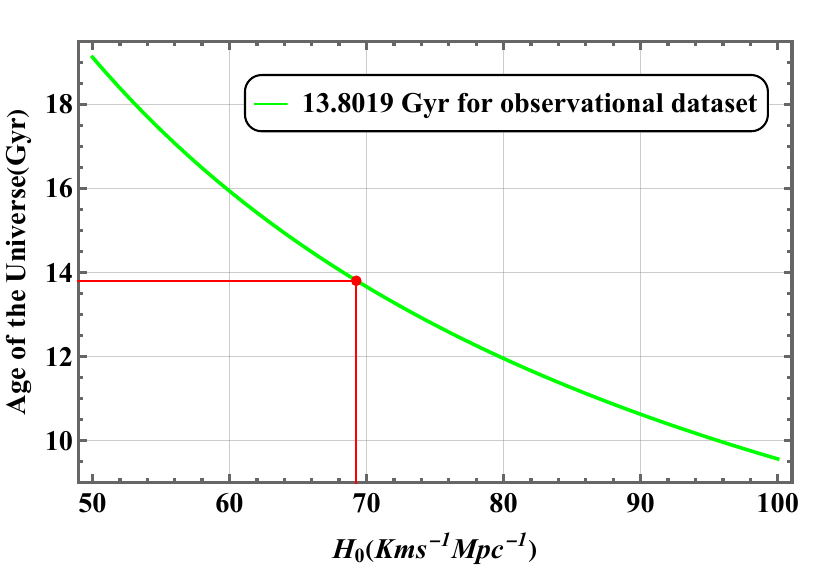}
\caption{{Plot for the age of the Universe. The model being considered effectively explains the evolution of the universe, as the obtained values align with the current bounds from Planck \cite{aghanim2020planck}.}}
\label{age}
\end{center}
\end{figure}

Fig.~\ref{age} shows the estimation of the Universe's age on the basis of the observational datasets considered. The analysis yields an age of approximately $t_0 = 13.8019$ Gyr. It may be noted that while calculating, we have considered the parameter values obtained in the earlier section. The Fig.~\ref{age} asserts that the age of the cosmos calculated using the reconstruction scheme described in previous sections is consistent with current observational limitations. The agreement between the model prediction and the observationally inferred age \cite{aghanim2020planck} validates the underlying cosmological assumptions associated with the reconstruction of THDE in $f(Q,C)$ gravity.

\subsection{Energy conditions}

In this section, we discuss different energy conditions, which are linear energy density and pressure combinations. The literature describes four types of energy conditions: weak energy condition (WEC), strong energy condition (SEC), null energy condition (NEC), and dominant energy condition (DEC). The ECs are the following in Table~\ref{EC} \cite{mandal2020,Bhagat2025exploring}:

\begin{table}[ht!]
\caption{The four classical energy conditions.}
\centering
\begin{tabular}{|c|c|c|}
\hline
\textbf{Energy Condition} &~~~~~~ \textbf{Inequalities} ~~~~~~&~~~~~ \textbf{In Terms of \( \omega = \frac{p}{\rho} \)}~~~~~ \\
\hline
NEC (Null Energy Condition) & \( \rho + p \geq 0 \) & \( \omega \geq -1 \) (if \( \rho > 0 \)) \\
\hline
WEC (Weak Energy Condition) & \( \rho \geq 0,\quad \rho + p \geq 0 \) & \( \omega \geq -1 \) \\
\hline
SEC (Strong Energy Condition) & \( \rho + p \geq 0,\quad \rho + 3p \geq 0 \) & \( \omega \geq -\frac{1}{3} \) \\
\hline
DEC (Dominant Energy Condition) & \( \rho \geq 0,\quad  \rho -p \geq 0 \) & \( \omega \leq 1 \) \\
\hline
\end{tabular}
\label{EC}
\end{table}
In Fig.~\ref{E}, we have illustrated the four energy conditions based on the reconstructed THDE in $f(Q,C)$ framework. This reconstruction has, as its natural consequence, a reconstructed Hubble parameter, which, after constraining with observations, has made us able to compute the $\rho_{total}$ by inserting the reconstructed $H(z)$ into THDE and thereafter computing $\rho_m$ accordingly. In Fig.~\ref{E}, the purple curve corresponds to $\rho_{total}$, which appears to be positive throughout. Combining this with other requirements for NEC, we have observed the validity of NEC, indicating $\omega\geq -1$, which is consistent with already obtained behavior of the EoS parameter. The blue curve indicates the NEC in Fig.~\ref{E}. The green curve representing SEC appears to be validated till $z\approxeq 0.75$. Later at lower redshifts, it crosses the boundary of $0$, satisfying the necessary condition for the accelerated expansion of the Universe. Finally, let us have a look at the yellow curve representing DEC. The behavior of DEC demonstrates that the reconstructed THDE satisfies the DEC throughout the cosmic evolution. To be more specific, we find that alongside the positive energy density, its value is greater than the absolute value of pressure in the different epochs of the evolution of the Universe. This behavior makes us infer that the effective energy momentum content that drives the late time acceleration of the Universe remains physically viable in our model constrained by observations and this model is capable of avoiding any exotic phenomenon like negative energy density. Furthermore, the validation of DEC complements the satisfaction of other energy conditions and collectively asserts the consistency of our model. Hence, we can infer that our reconstructed THDE model in the $f(Q,C)$ framework provides a viable explanation for the accelerated expansion of the late time alongside adhering to the energy conditions suitable for physically viable cosmology.
\begin{figure}
\begin{center}
\includegraphics[height=3.5in]{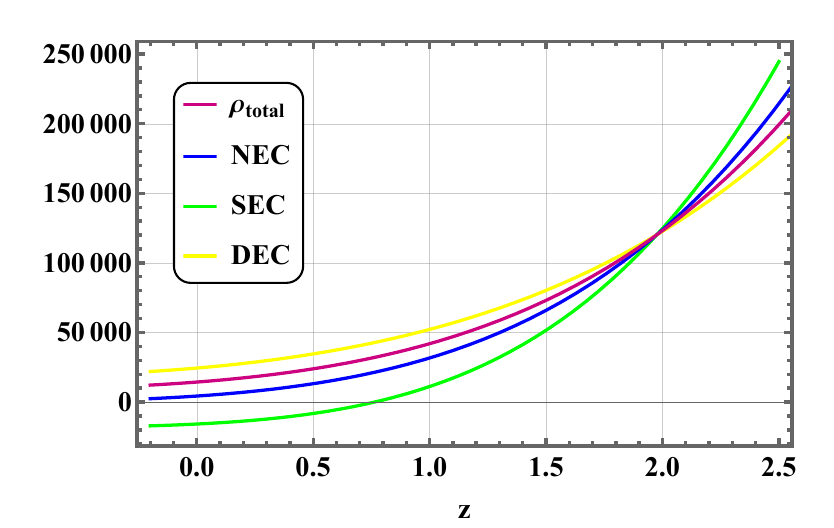}
\caption{{Pictorial presentation of the energy conditions plotted against redshift.}}
\label{E}
\end{center}
\end{figure}

\section{Concluding remarks}
In the research work reported in the previous sections, we have explored the cosmology of reconstructed THDE in a modified gravity framework namely $f(Q,C)$, with the motivation of combining entropy-based dark energy models with geometrically motivated modified gravity to explain late-time cosmic acceleration. It was illustrated that the reconstructed model exhibits significant sensitivity to the parameter space $(H_0,  a_0, n, \delta, \zeta,r_d)$ and initial conditions, with the evolution of the EoS parameter and other cosmological quantities highly dependent on these parameters. {\color{black} A comprehensive MCMC analysis using observational datasets (CC, Pantheon$^{+}$ and DESI DR2) was performed (see Figure~\ref{T}), yielding best-fit values ($H_0 = 69.25\pm0.90$ km/s/Mpc, $a_0=9.9\pm1.6$, $n = 0.530\pm0.090$, $\delta = 1.6\pm1.4$, $\zeta = 2.5\pm1.7$ \& $r_d=146.5\pm1.3$) that demonstrate strong consistency with the combined observational datasets. {The obtained sound horizon $r_d$ is consistent with the $\Lambda$CDM CMB prediction $r_d \approx 147~\mathrm{Mpc}$. The resulting value $H_0 r_d \approx 1.01 \times 10^4~\mathrm{km\,s^{-1}}$ is also consistent with the BAO constraint $H_0 r_d \approx 9.9 \times 10^3~\mathrm{km\,s^{-1}}$ \cite{karim2025desi,rd}, supporting the compatibility of the reconstructed THDE-$f(Q,C)$ model with early-universe physics. Although the difference in the minimum $\chi^2$ values, $\Delta\chi^2_{\min}=2.041$, slightly favors the $\Lambda$CDM model, the reconstructed THDE–$f(Q,C)$ framework yields a fit of comparable quality, demonstrating that the model remains consistent with current observational constraints (see Tables ~\ref{OD} and \ref{OD1}). Figure~\ref{T1} shows the running mean convergence of the MCMC chains for the sampled cosmological parameters. After an initial unstable phase, the running mean stabilizes around nearly constant values, indicating that the chains have reached a stationary distribution and confirming convergence \cite{Sampling}.} Overall, Table~\ref{OD} demonstrates that the proposed framework provides a good fit to the observational data and supports non-extensive holographic dark energy as a viable extension of the standard cosmological model. The theoretical curves of the constants $C_1$, $C_2$, and $C_3$ in Fig.~\ref{Diff} exhibit a smooth, monotonically increasing trend with redshift, corresponding to a decrease of $H(z)$ as the Universe evolves. While different choices of the integration constants preserve the qualitative behavior of the model, the associated uncertainty band broadens from $C_1$ to $C_3$, reflecting the sensitivity of the reconstruction to this parameter. Nevertheless, the strong overlap with the $\Lambda$CDM prediction across the entire redshift range demonstrates the continued observational viability of the model considered. The reconstructed THDE model closely aligns with the $\Lambda$CDM model throughout the redshift range remaining within observational uncertainties (Fig.~\ref{H}), suggesting it can effectively describe late-time cosmic acceleration without strictly adhering to $\Lambda$CDM assumptions. It is noteworthy that the widths of the confidence intervals have significantly reduced from higher to lower redshift and at lower redshifts, $1\sigma$ and $2\sigma$ confidence intervals are almost coincident. The reconstructed model (Fig.~\ref{P}) exhibits close agreement with the Pantheon$^{+}$ dataset over the entire redshift range. The closeness implies statistical consistency with the observational data.}

The reconstructed $f(Q,C)$ was presented as a function of cosmic time $t$. The reconstructed function $f(t)$ reflected the dynamics of THDE inside the modified gravity framework $f(Q,C)$, showing its dependency on major cosmological factors via higher-order derivatives. Limiting cases illustrate the impact of $n$ and $\delta$ on $f(t)$: for $n \to 0$, the power-law terms dominate, while for $t \to \infty$, the behavior of $f(t)$ is dominated by the sign of the exponent, resulting in decay or growth. Initially, $t^{-2}$ and $t^{-4}$ provide the most significant contributions. This study shows how the model may capture the intricate evolution of the Universe at various cosmic stages. Statefinder diagnostic, through the evolutionary trajectory of the pair $(r, s)$ (Fig.~\ref{rs}), shows that the model passes through the $\Lambda$CDM fixed point which strongly indicates that the THDE model reconstructed in \(f(Q,C)\) gravity revolves around the \(\Lambda\)CDM phase of the Universe. The analysis of the evolutionary trajectory in the $(r, q)$ Statefinder plane, as depicted in Fig.~\ref{rq}, revealed some features of the THDE model reconstructed within the $f(Q,C)$ gravity framework. The trajectory originated from the standard matter-dominated era (decelerated) $(q = 0.5, r = 1)$ and asymptotically approaches the late-time accelerated phase $(q = -1, r = 1)$, passing through the transition region. The presence of the $\Lambda$CDM point $(r = 1)$ along the trajectory further demonstrates the compatibility of the reconstructed model with the standard cosmological paradigm at late times, while still allowing room for deviations in the early Universe. The transition of the deceleration parameter $q(z)$ along with its $1\sigma$ and $2\sigma$ confidence bands (Fig~\ref{q}) from positive to negative values indicates the transition from decelerated to accelerated phase of the Universe, highlighting the viability of the THDE model reconstructed in $f(Q,C)$ gravity as a consistent and observationally supported model to study the late-time acceleration of the Universe. In the early Universe, the EoS parameter \( {\omega_{tot}(z)} \)} along with its $1\sigma$ and $2\sigma$ confidence intervals (Fig.~\ref{w}) was positive, leading to various expansion behaviors. In the context of THDE in \( f(Q,C) \) gravity, the reconstructed EoS evolves into the negative regime, showing quintessence-like behavior (\(-1< {\omega_{tot}(z)} < -\frac{1}{3} \)) without crossing the phantom boundary, approaching \( {\omega_{tot}(z)} \approx -1 \) at late times. 

{\color{black} The redshift evolution of the jerk parameter $j(z)$ (Fig.~\ref{jerk}) is compared with the $\Lambda$CDM benchmark $j=1$. While mild deviations may appear at intermediate redshifts, the model approaches $j=1$ at late times and remains consistent with the combined observational datasets (CC+Pantheon$^{+}$+DESI DR2). The redshift evolution of the snap parameter $s(z)$ with $1\sigma$ and $2\sigma$ confidence regions is shown in Fig.~\ref{snap} and compared with the $\Lambda$CDM prediction. The obtained present value $S_0=-0.428^{+0.075}_{-0.064}$ is close to the $\Lambda$CDM estimate $S_0\approx -0.35$ for $\Omega_{m_0}=0.3$ \cite{Jesus}. The discussion is restricted to the observed redshift range, while $z<0$ only represents the theoretical future evolution of the model.} The current values of the cosmological parameters constrained using the combined datasets CC+Pantheon$^{+}$+DESI DR2 for reconstructed THDE in $f(Q,C)$ gravity are presented in Table \ref{CV}. The estimated age of the Universe (Figure~\ref{age}), approximately $13.8019~ Gyr$, remains consistent with current observational constraints, further validating the model's assumptions. The agreement of the reconstructed age with current observational constraints indicates the validity of the cosmological assumptions that underpin the THDE model in the $f(Q,C)$ gravitational framework, bolstering the dependability of our reconstruction approach. In the last phase of the study, a rigorous analysis is presented for the energy conditions. In summary, the energy condition analysis for the reconstructed THDE model within the $f(Q,C)$ gravitational framework, as represented in Fig.~\ref{E}, demonstrates that the total energy density $\rho_{\text{total}}$ remains positive throughout cosmic history and meets the Null Energy Condition (NEC). The Strong Energy Condition (SEC) holds up to $z \approx 0.75$ before violating at lower redshifts, indicating the start of accelerated expansion. The Dominant Energy Condition (DEC) remains constant across all epochs, ensuring a physically viable energy-momentum content devoid of unusual features. The consistent meeting of all important energy conditions strengthens our model. As a result, the reconstructed THDE in $f(Q,C)$ gravity emerges as a convincing and empirically consistent alternative to orthodox cosmology, providing a plausible explanation for the Universe's late-time acceleration.

{\color{black} In this study, we have reconstructed a modified gravity model of the form $f(Q,C)$ supplemented by THDE and examined. Furthermore, we examined its viability as a late-time extension of $\Lambda$CDM in the context of precision cosmology. The framework, explored in this study, is motivated by the persistence of cosmological tensions that increasingly challenge the standard paradigm. The $f(Q,C)+$THDE construction introduces late-time deviations through the non-metricity scalar $Q$, an explicit matter creation term $C$, and a non-extensive holographic dark energy sector. This study is motivated by the persistence of the statistically significant cosmological tensions that have been highlighted in the recent comprehensive reviews of precision cosmology \cite{DiValentino2025CosmoVerse,DiValentino2025Tensions}.The reconstructed background solutions admit mild, redshift-dependent modifications of the expansion rate, qualitatively consistent with the localized expansion anomalies inferred from DESI-DR2 BAO and supernova observations \cite{Mukherjee2025Expansion,Lodha2025DESI}.  Moreover, this reconstructed model of $f(Q,C)$ is qualitatively consistent with recent indications of localized expansion-rate anomalies from DESI and supernova data. Within the parameter space explored in this study, the model does not exhibit instabilities as reflected by the confrontation with the recent observational data and admits a smooth transition from matter domination to accelerated expansion. In this reconstruction approach the effective dark energy component has the scope of emerging dynamically at low redshifts, while leaving early-universe observables largely unaffected. This separation of early- and late-time physics is a central requirement identified for viable tension-mitigating models \cite{DiValentino2025CosmoVerse,Casertano2025LDN}. In summary, the $f(Q,C)+$THDE model constitutes a theoretically consistent framework for exploring late-time departures from $\Lambda$CDM in the era of precision cosmology. Although a further definitive assessment of its potential to resolve specific tensions requires full joint likelihood analyses, the present results can make us consider this model within the class of well-motivated extensions identified as promising in recent cosmological studies \cite{DiValentino2025Tensions}.}

While concluding, let us have some comparison with the existing literature. In comparison to previous studies, our reconstructed THDE model within the $f(Q,C)$ gravity framework offers a compelling alternative to traditional cosmological models. While Saridakis \cite{saridakis2021} explored the dynamics of dark energy and modified gravity, our approach uniquely combines entropy-based dark energy models with geometrically motivated modifications to gravity, capturing a richer parameter space. Our reconstructed THDE model in the $f(Q,C)$ gravity framework led to results that were compatible with observational data and aligned with the studies of ~\cite{Sultana2024,Shekh,Oikonomou}, where modified gravity frameworks like $f(R,T)$, $f(Q)$, and $f(G)$ were used to explain cosmic acceleration and dark energy phenomena. Another work relevant to our study, \cite{Oikonomou} demonstrated that reconstructed models with appropriately chosen parameter spaces can mimic $\Lambda$CDM behavior at late times while allowing for deviations at early epochs, thus accommodating both observational consistency and theoretical flexibility. Future research could look into how higher-order gravity modifications interact with other entropy-based dark energy models, widening the parameter space and improving predictions for cosmic acceleration and structure formation. {Furthermore, in reference to the jerk parameter, we proposed a detailed assessment of the model consistency with the expansion history of the universe at very high redshit $z\approx1100$ as future work.}

{\color{black}\subsection*{Acknowledgments}
The insightful comments of the anonymous reviewer are gratefully acknowledged. The authors also acknowledge the Inter-University Centre for Astronomy and Astrophysics (IUCAA), Pune, India, for providing library and computing facilities during their scientific visits in 2024 and 2025, which supported the completion of this work.}

\section*{CRediT Author Contribution Statement}

\noindent \textbf{Sanjeeda Sultana:} Writing – original draft, Methodology, Investigation, Conceptualization. \\
\textbf{Surajit Chattopadhyay:} Writing – review \& editing, Supervision, Methodology, Conceptualization.

\section*{Declaration of Competing Interest}

\noindent The authors hereby declare that they have no conflict of interest with anyone concerning the paper.

\section*{Declaration Regarding AI Tool Use}

\noindent The authors hereby declare that no AI tools were used in the generation of the research data or findings. However, AI tools such as QuillBot and Grammarly were employed solely for language improvement and grammar correction in the manuscript.

{\color{black}\section*{Data Availability Statement}
\noindent For the analysis reported in the paper we have utilized the data available in \cite{CC4,karim2025desi,scolnic2022pantheonplus}.}

\end{document}